\documentclass[aps,prr,superscriptaddress,twocolumn]{revtex4-2}
\usepackage{dcolumn}
\usepackage{graphicx} 
\usepackage{amsmath}
\usepackage{bm}
\usepackage{xcolor}
\usepackage[T1]{fontenc}
\usepackage[utf8]{inputenc}
\usepackage{enumerate}
\usepackage{physics}
\usepackage{amssymb}
\usepackage{enumitem} 

\begin{document}
	
\title{Ferroelectricity in dipolar liquids: the role of annealed positional disorder}
	
\author{Maria Grazia Izzo}
\affiliation{Department of Molecular Sciences and Nanosystems, Ca' Foscari University of Venice, Via Torino 155, 30172 Venezia Mestre, Italy}
\email{mariagrazia.izzo@unive.it}
	
	\date{\today}
	
	\begin{abstract}
Ferroelectric ordering in polar liquids has been observed in numerical simulations and liquid-crystal experiments. Within mean-field framework, this behaviour remains associated with sample-shape dependent, surface contribution to the free energy, which does not vanish in the thermodynamic limit due to the long-range nature of dipolar interaction. 
Yet, numerical simulations performed under conducting periodic boundary conditions, for which the surface contribution vanishes, still exhibit ferroelectric order, pointing to an intrinsic bulk origin of the transition. 
Moving beyond the mean-field approximation, Kirkwood’s seminal study of the dielectric properties of polar liquids emphasized the role of hindered dipolar rotation in shaping the corresponding pair correlations. In Kirkwood's analysis, hindered rotation stems from the mean force between nearest-neighbor dipoles, placing the focus on local structure.
Introducing a different perspective while retaining the central role of hindered dipolar rotation in the onset of ferroelectricity, the present study establishes, as an original finding, that annealed averaging of dipolar interaction over positional disorder generates hindered dipolar rotation favoring dipole alignment, and able to drive a ferroelectric phase transition. 
As a result, unlike approaches centered on local structure, ferroelectricity emerges not in spite of the liquid nature, but because of it. This ferroelectric phase transition is shown to be intrinsic to the bulk. 
Annealed averaging over positional disorder defines an effective dipolar interaction that is shorter-ranged than the bare potential. This is analogous to the Keesom interaction where screening arises from annealed dipolar disorder.
Derived within classical density functional theory, these findings are exact for dimensions $d \to \infty$ and remain valid within the optimized cluster expansion for $d \geq 3$.
	\end{abstract}
	
\maketitle
\section{Introduction} \label{intro}
\begin{figure}[t]
	\centering
	\includegraphics[width=0.9\linewidth]{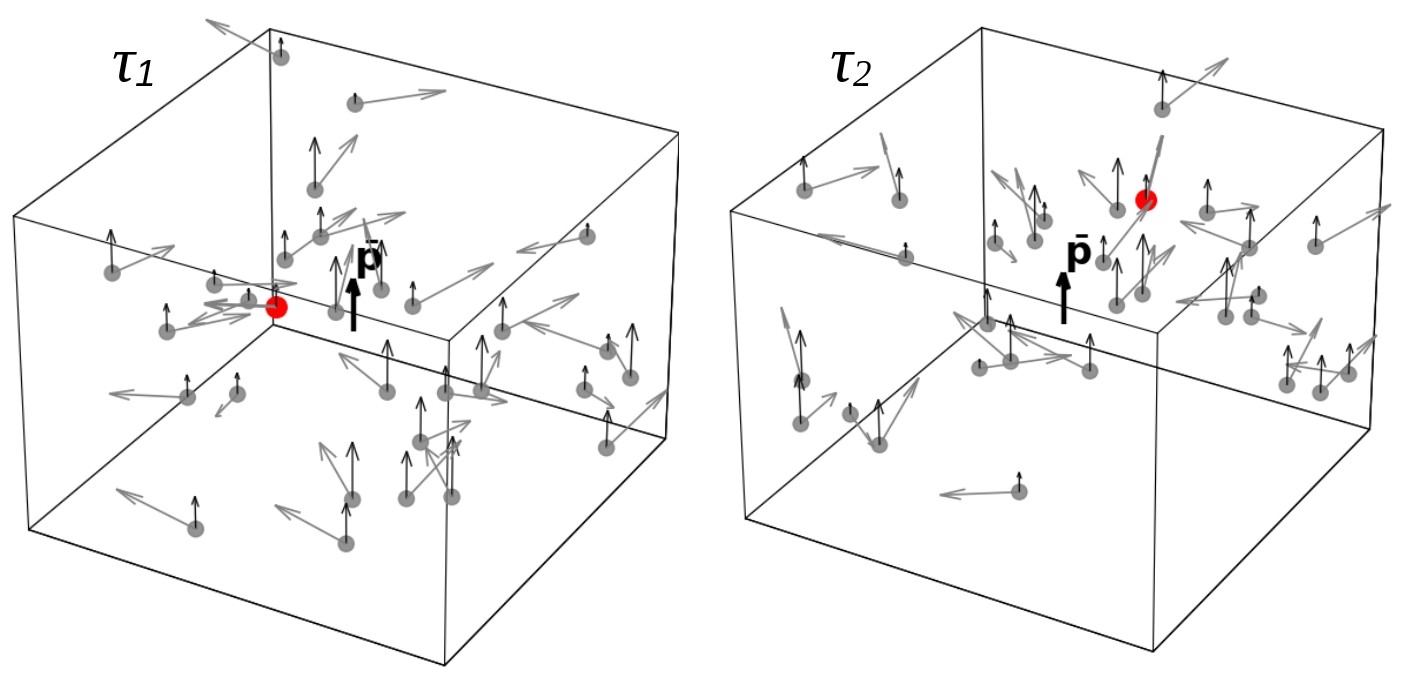}
	\caption{Schematic representation of two configurations of a dipolar liquid in three dimensions at times $\tau_{1(2)}$ in the ferroelectric phase. Grey arrows show the orientations of the particle dipole moments, thin black lines represent their projections along the direction of the macroscopic polarization per particle, shown by the thick black arrow at the center of the simulation box. Simple ferroelectric order, set only by a macroscopic polarization, does not constrain translational degrees of freedom and the particle highlighted in red is free to move in space, independently of the orientation of the dipole that it carries.}
	\label{fig1}
\end{figure}

The possibility of a ferroelectric phase transition in dipolar liquids traces back to the studies of Debye, Onsager, and Kirkwood \cite{Debye, Onsager,Kirkwood}. Following numerical simulations of dipolar liquids have reported transitions toward dipole‐ordered states \cite{Wei,Weis1,Weis2,Bartke,Camp,Levesque}, while recent experiments in liquid‐crystals have provided evidence for a ferroelectric nematic phase \cite{Chen, Nishikawa, Martelj}. A renewed interest in the topic is further supported by recent findings showing that supercooled water in its low-density phase exhibits properties consistent with a ferroelectric phase \cite{Izzo, Malosso}. In Ref. \cite{Izzo} it is furthermore shown how the liquid-liquid phase transition in supercooled water may be driven by a ferroelectric phase transition. 
A ferroelectric phase transition is characterized by the spontaneous emergence of a macroscopic polarization below a critical value of suitable thermodynamic control parameters, such as temperature or pressure.
For liquids as well as solids, the order parameter of a ferroelectric phase transition is the macroscopic polarization vector per particle, $\bar{\mathbf p}$. In the case of a polar liquid of $N$ particles labeled by $i=1,\ldots,N$, each carrying a dipole moment $p\hat{d}_i$, it is defined as
\begin{eqnarray}
	\bar{\mathbf{p}}= \langle \lim_{N \rightarrow \infty}\frac{1}{N} \ p \  \sum_{i=1}^{N} \hat{d}_i \rangle= \nonumber \\ \lim_{T\rightarrow \infty} \frac{1}{T} \sum_{\tau=0}^{T} \lim_{N\rightarrow \infty}  \frac{1}{N}\ p  \sum_{i=1}^{N} \hat{d}_i(\tau), \label{P}
\end{eqnarray} 
where $p$ is the magnitude of the rigid particle dipole moment, $\langle \ \rangle$ denotes the ensemble average, $\tau$ is the time variable, and ergodicity allows one to set the equality in Eq. \ref{P}. In the paraelectric phase $\bar{\mathbf{p}}=0$, whereas in the ferroelectric phase $\bar{\mathbf{p}} \neq 0$, corresponding to the spontaneous breaking of continuous rotational symmetry. In a solid, where dipoles occupy fixed lattice sites, microscopic configurations corresponding to a ferroelectric phase are readily identified. The dipoles align so that the lattice-averaged polarization remains nonzero and essentially configuration independent, yielding a finite ensemble average.
In a liquid, an analogous picture emerges once the lattice constraint is removed and individual dipoles are free to move in space. A ferroelectric liquid phase is then characterized by a nonzero single-particle dipole moment averaged over all particles, whose value remains approximately configuration independent, as in the solid. A schematic illustration is shown in Fig. \ref{fig1}.
However, while solids possess a frozen lattice, so that only dipolar dynamics must be constrained to maintain ferroelectric order, in liquids the emergence of more complex ferroelectric ordering, such as chiral order, can also constrain translational motion in addition to dipolar rotations. 
A nonzero macroscopic polarization alone does not establish a ferroelectric phase transition. In the thermodynamic limit, the latter is characterized by non-analytic behavior of the free energy or its derivatives with respect to a control parameter, as prescribed by the Ehrenfest classification \cite{Landau}. If a phase transition is intrinsic to the bulk, its existence and critical behavior must be independent of boundary conditions, surface terms, and sample shape. This becomes especially relevant for long-range interactions, such as dipolar forces, as discussed below.

\begin{figure}[t]
	\centering
	\includegraphics[width=0.7\linewidth]{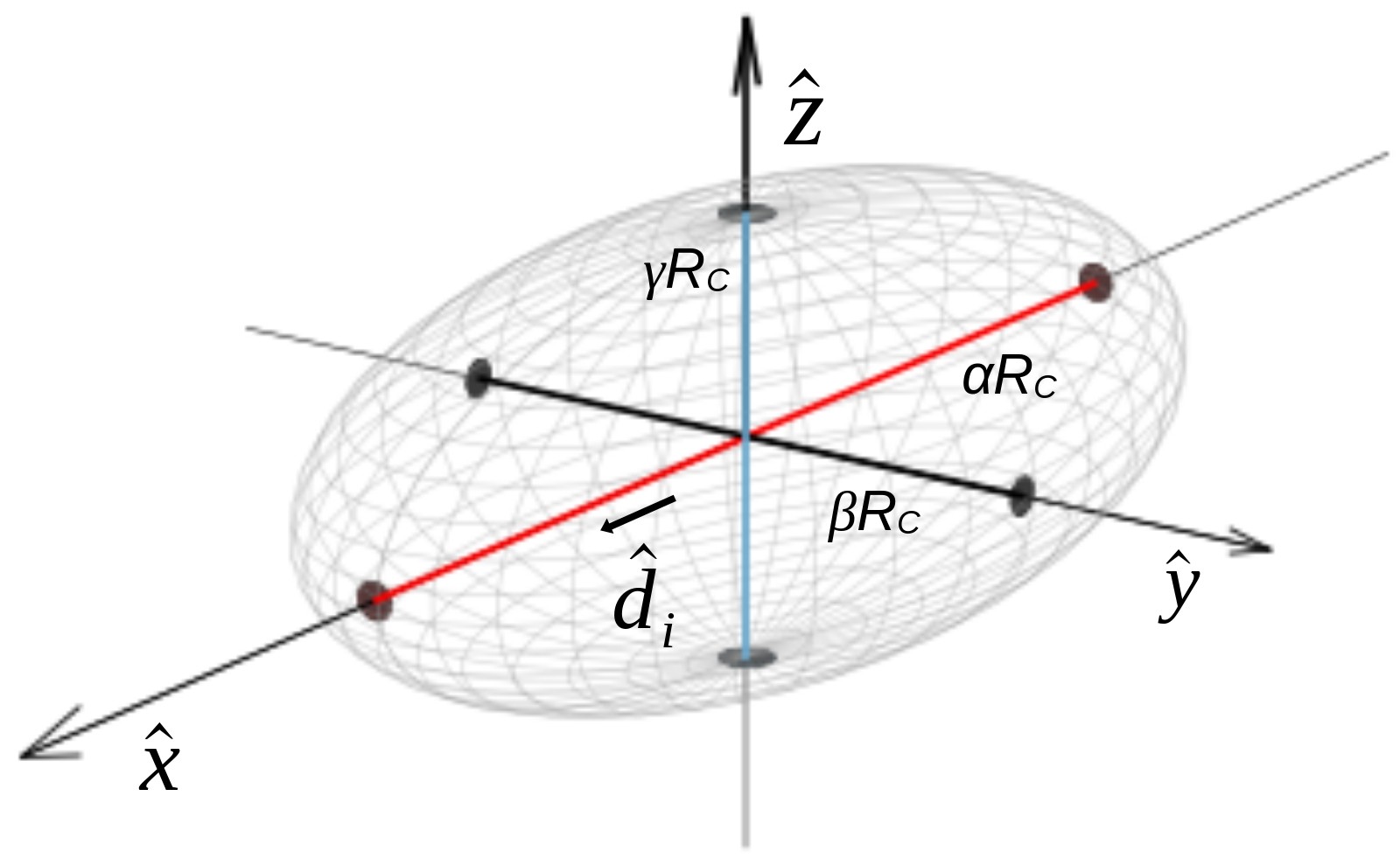}
	\caption{Sketch of the ellipsoidal sample used to estimate, via octahedral quadrature, the surface contribution to the excess free energy associated with the dipolar potential in a spatially homogeneous system within the mean-field approximation.}
	\label{fig2}
\end{figure}

Classical density functional theory (DFT) provides a formally exact framework in which the free energy, $F$, of a many-body system is expressed as a functional of the one-body density field, $\tilde{\rho}$ \cite{Hansen}. The equilibrium function $\tilde{\rho}$ can be obtained by applying a variational principle to $F[\tilde{\rho}]$. Owing to its unbiased formulation, classical DFT is a powerful tool for the study of phase transitions.
Moreover, it applies to both crystalline \cite{Rick, Groh1, Ding} and liquid phases and, when combined with replica theory \cite{Parisi}, can be further extended to glassy states \cite{Klapp}. A common DFT scheme consists in decomposing the pair potential, $v$, into a reference part, $v_0$, the pair potential acting in the reference system, and a perturbative part, $w_p$ \cite{Hansen}. In the case of dipolar liquids, $v_0$ is the hard-sphere or the Lennard-Jones potential, and $w_p$ is the dipolar interaction.
The family of intermediate potentials
\begin{eqnarray}
	v_{\lambda}(r,\hat{r}_{ij},\hat{d}_i,\hat{d}_j)=v_0(r,\hat{r}_{ij})+\lambda w_p(r,\hat{r}_{ij},\hat{d}_i,\hat{d}_j),  \\ 0 \leq \lambda \leq 1 \label{v_lambda},\nonumber
\end{eqnarray}
is then introduced to parametrize an adiabatic path from the reference system ($\lambda=0$) to the fully interacting system ($\lambda=1$) \cite{Hansen}. In Eq. \ref{v_lambda}, $\hat{d}$ is the dipole unit vector, subscripts $i$ and $j$ indicate association with particle $i$ or $j$, respectively, $\mathbf{r}$ is a space vector and $r=|\textbf{r}_i-\textbf{r}_j|$, $\hat{r}_{ij}=\frac{\textbf{r}_i-\textbf{r}_j}{r}$. The free energy functional, in its more general DFT formulation, is
\begin{eqnarray}
	F[\tilde{\rho}]=F_{v_0}[\tilde{\rho}]+\mathcal{F}[\tilde{\rho}], \label{F_DFT0}
\end{eqnarray}
where $F_{v_0}[\tilde{\rho}]$ is the free energy of the reference system and $\mathcal{F}$ is the excess free energy associated to $w_p$ \cite{Hansen},
\begin{align}
	\mathcal{F}[\tilde{\rho}]
	&=
	\frac{1}{2}
	\int_0^1 d\lambda
	\int d\textbf{r}_i d\textbf{r}_j
	d\hat{d}_i d\hat{d}_j\,
	\tilde{\rho}(\textbf{r}_i,\hat{d}_i)
	\nonumber\\
	&\quad \times
	g^{(2)}_{\lambda}
	(\textbf{r}_i,\textbf{r}_j,\hat{d}_i,\hat{d}_j)
	\tilde{\rho}(\textbf{r}_j,\hat{d}_j)
	w_p(r,\hat{r}_{ij},\hat{d}_i,\hat{d}_j).
	\label{F_DFT}
\end{align}
whit $d  \hat{d}$ being the differential solid angle element.
In Eq. \ref{F_DFT}, $g^{(2)}_{\lambda}(\textbf{r}_i,\textbf{r}_j,\hat{d}_i, \hat{d}_j)$ is the pair correlation function of the full system governed by the potential $v_{\lambda}(r,\hat{r}_{ij},\hat{d}_i,\hat{d}_j)$. 
The one-particle density field for dipolar liquids is
\begin{eqnarray}
	\tilde{\rho}(\textbf{r},\hat{d})=\sum_{i=1}^N \delta (\hat{d}-\hat{d}_i)\delta (\mathbf{r}-\mathbf{r}_i)=\rho (\textbf{r}) \zeta (\mathbf{r},\hat{d}),	 \label{rho_tilde}
\end{eqnarray}
where $\rho(\textbf{r})$ and $\zeta (\textbf{r}, \hat{d})$ are respectively the particle number density marginalized over dipole orientation and the probability distribution of dipole orientation at $\textbf{r}$. $\delta( \ )$ is the Dirac delta function.
As appropriate for a liquid, spatial homogeneity is assumed, making $\tilde{\rho}(\textbf{r}, \hat{d})$ independent of $\textbf{r}$. However, no assumption is imposed on the dipole orientational probability distribution, thereby allowing for possible dipolar ordering within the liquid. $\tilde{\rho}$ then factorizes as 
\begin{eqnarray}
	\tilde{\rho}=\rho \ \zeta (\hat{d}) \label{rho_om},
\end{eqnarray}
where $\rho$ is the particle number density. In a DFT scheme the physical insight and technical challenges are entirely encoded in the characterization of $g^{(2)}_{\lambda}$. The lowest level of approximation is a mean-field theory, where $g^{(2)}_{\lambda}=1$. The excess free-energy functional for the one-particle density in Eq. \ref{rho_om} under this approximation becomes
\begin{eqnarray}
	\mathcal{F}^{MF}[\tilde{\rho}]=\frac{1}{2} N\rho\int d\hat{r}_{ij}r^2 dr d  \hat{d}_i d  \hat{d}_j  w_p(r, \hat{r}_{ij},\hat{d}_i, \hat{d}_j).	 \label{F_DFT_MF}
\end{eqnarray} 
If one considers the bare dipolar interaction potential, e.g. in three-dimensional space,
\begin{eqnarray}
	w_p^{(3)}(r, \hat{r}_{ij},\hat{d}_i, \hat{d}_j)=-p^2 \frac{1}{r^{3}} [3(\hat{d}_i \cdot \hat{r}_{ij})(\hat{d}_j \cdot \hat{r}_{ij})-\hat{d}_i\cdot \hat{d}_j],	 \label{w_p0}
\end{eqnarray} 
the associated integral in Eq. \ref{F_DFT_MF} exhibits conditional convergence \cite{Wei, Theis, Osipov, Osipov1, Stenqvist}. Its value depends on the order of integration over the radial and angular variables, $r$ and $\hat{\mathbf r}_{ij}$, both in the limits of short- and long-range interparticle separations, and therefore on the shape of the integration domain in the thermodynamic limit. To avoid conditional convergence at short $r$, a physically motivated way is to take $w_p$ vanishes inside the core region of $v_0$, where it is indeed ineffective due to the hard-core repulsion.
Alternatively, allowing for an arbitrary form of $w_p$ within the core region, the spatial integral of the dipolar interaction may be evaluated over the region outside a spherical cavity, with the cavity radius then taken to zero \cite{Osipov}. However, it has been shown that the resulting free energy depends on the arbitrary choice in which the perturbing potential is defined within the core region \cite{Hoye}.
This approach therefore will not be considered further. Both the Debye and Onsager models, as well as the Wertheim mean-spherical approximation \cite{Wertheim}, can be retrieved within this mean-field scheme, each corresponding to a specific assignment of the perturbing potential inside the core \cite{Hoye}. Due to the conditional convergence at long interparticle distances, the mean-field contribution to the excess free energy depends on the macroscopic shape of the sample \cite{Groh}. An intuitive picture emerges by considering an ellipsoidal sample with semiaxes $\alpha R_c$, $\beta R_c$, and $\gamma R_c$ along $\hat{x}$, $\hat{y}$, and $\hat{z}$, respectively (Fig.~\ref{fig2}), and approximating the surface term in Eq.~\ref{F_DFT_MF} associated with the ellipsoidal geometry, $\mathcal{F}_{S_e}^{MF}$, by octahedral quadrature \cite{Stroud}. The octahedral quadrature approximates angular integrals over three-dimensional space by a discrete sum over the six Cartesian directions, each with weight $2\pi/3$. Within this approximation,     
\begin{eqnarray}
	\mathcal{F}_{S_e}^{MF}(R_c)=\frac{1}{2}N\rho\frac{2 \pi}{3}\sum_{\hat{r}_k=\pm\hat{x},\pm\hat{y},\pm\hat{z}} w_p^{(3)}(1,\hat{r}_k)\ln\big(R(\hat{r}_k)\big),
\end{eqnarray}
where $\hat{x},\hat{y},\hat{z}$ are the Cartesian directions and $R(\hat{r}_k)$ is the distance from the center to the surface of the ellipsoid along direction $\hat{r}_k$, i.e., $R(\pm\hat{x})=\alpha R_c$, $R(\pm\hat{y})=\beta R_c$, and $R(\pm\hat{z})=\gamma R_c$.  
Taking $\hat{d}_i=\hat{x}$, in the macroscopic limit $R_c \rightarrow \infty$, one obtains
\begin{eqnarray}
	\mathcal{F}_{S_e}^{MF}=-\frac{1}{2}N\rho p^2 \frac{2 \pi}{3} \hat{d}_i \cdot \hat{d}_j [4 \ln(\alpha)-2 \ln(\beta)- 2\ln(\gamma)]. \nonumber  \\
\end{eqnarray}
The surface term is thus different from zero and shape dependent. If the coefficient in the brackets is positive, the bulk system exhibits a tendency toward dipole alignment, if negative it favors antiparallel dipole alignment. When the ellipsoid reduces to a sphere, $\alpha=\beta=\gamma$, and $\mathcal{F}_{S_e}^{MF}$ vanishes. It is worth emphasizing that, for a bulk homogeneous and thus isotropic system, the bulk contribution to the integral in Eq. \ref{F_DFT_MF} vanishes. Indeed, isotropy translates in a uniform distribution of $\hat{r}_{ij}$ over the solid angle, so that integration over $\hat{r}_{ij}$ reduces to an average over the solid angle. In particular,
$\langle(\hat{d}_i \cdot \hat{r}_{ij})(\hat{d}_j \cdot \hat{r}_{ij})\rangle = \frac{1}{d}\hat{d}_i \cdot \hat{d}_j$, where $d$ is the spatial dimension. This follows directly from rotational
invariance, which implies $\langle \hat r_{ij}^{\alpha}\hat r_{ij}^{\beta} \rangle = \frac{1}{d}\delta_{\alpha\beta}$,
where $\alpha$ and $\beta$ denote the components of the unit vector
$\hat r_{ij}$.
This offsets the remaining term of the dipolar potential, thus making the bulk contribution to the integral in Eq. \ref{F_DFT_MF} vanish. For a homogeneous system with one-particle density given in Eq. \ref{rho_om}, $\mathcal{F}^{MF}$ thus reduces to a pure surface contribution. Note that the conditional convergence of real-space integrals of the dipolar potential persists even in the limit $d \rightarrow \infty$. Most theoretical studies predicting the occurrence of a ferroelectric phase transition in dipolar liquids rely on mean-field approximation, or assume the mean-field contribution as the leading term in the free energy \cite{Osipov, Osipov1, Cattes, Groh}. The resulting ferroelectric phase transition cannot thus be regarded as a genuine bulk phase transition.
The first question addressed in this study is the following: is this conclusion merely a consequence of the mean-field approximation? Recast in more general terms:
\begin{enumerate}[label=(\roman*), series=ferro]
	\item Does a ferroelectric phase transition exist in dipolar liquids as an intrinsic bulk property? 
\end{enumerate}
This issue is far from being only conceptual. 
Consider, for instance, water. The proposed liquid–liquid phase transition in the supercooled regime has been invoked to explain the thermodynamic anomalies of bulk water. Ref. \cite{Izzo} showed that, insofar as water can be described as a dipolar liquid, such a transition may be driven by ferroelectric ordering. Since both the liquid–liquid transition and the associated anomalies are intrinsic bulk properties, consistency requires the underlying ferroelectric transition to be intrinsic to the bulk.
Were this not the case, one would need to investigate the role of interactions other than dipolar in accounting for the ferroelectric order observed in the low-density phase of supercooled water in numerical simulations \cite{Izzo, Malosso}. 
Such interactions may nonetheless influence nonuniversal features of the transition. However, if dipolar interactions are not the primary driving mechanism, the criteria for identifying such alternative interactions and their impact would be different.
The surface contribution to the extra free energy in Eq. \ref{F_DFT_MF} vanishes when the liquid is embedded in a conducting medium, as well as in numerical simulations under Ewald summation with conducting periodic boundary conditions.
Yet numerical simulations consistently report ferroelectric ordering in dipolar liquids under these conditions \cite{Wei,Weis1,Weis2,Camp}, including those analyzed in Ref. \cite{Izzo}, providing support for a positive answer to the questions raised above.
However, since identifying a genuine phase transition in finite-size simulations remains nontrivial, a theoretical understanding is desirable.

In dipolar liquids without ferroelectric order, free rotational motion of the dipoles screens dipolar interactions \cite{Mercer, Frodl}, and makes the bare long-range dipolar potential shorter ranged. This gives rise to the so-called Keesom interaction \cite{Keesom}. 
Conversely, whether the particles translational motion can induce a screening of the dipolar interaction has been overlooked. In a liquid, particle positions are not quenched but fully equilibrated and treated as statistical variables in the equilibrium ensemble. In the following, this property will be referred to as annealed positional disorder. As a consequence, the free energy, up to the ideal-gas contribution, is expressed in terms of the partition function $Z$ as
\begin{eqnarray}
	F=-\frac{1}{\beta} \ln Z =\nonumber \\ -\frac{1}{\beta} \ln \langle e^{-\beta \sum_{i<j} [v_0(r_{ij})+w_p(r_{ij},\hat{r}_{ij},\hat{d}_i,\hat{d}_j)]} \rangle_{\{\mathbf{r}_i,\hat{d}_i\}} \label{F_liq}, 
\end{eqnarray}
where $\beta=\frac{1}{k_BT}$, $T$ is the temperature and $k_B$ the Boltzmann constant. The explicit dependence of $r$ on the particle indices has been introduced for clarity, the case $\lambda=1$ of the family of potentials in Eq. \ref{v_lambda} is considered, and $\langle \ \rangle_{\{\mathbf{r}_i,\hat{d}_i\}}$ denotes the equilibrium ensemble average over all configurations of particle positions $\mathbf{r}_i$ and dipole orientations $\hat{d}_i$. Since the system is assumed to remain liquid in its positional degrees of freedom even upon the onset of dipolar order, the distribution of $\hat{r}_{ij}$ is always isotropic within the present framework.
The screening of the dipolar interaction by translational degrees of freedom can then be understood as arising from an annealed average over $\hat{r}_{ij}$. 
Given a two-body potential interaction $W(r_{ij},\hat{r}_{ij},\hat{d}_i,\hat{d}_j)$, its corresponding annealed average over $\hat{r}_{ij}$, $\bar{W}(r_{ij},\hat{d}_i,\hat{d}_j)$ is defined through $e^{-\beta \bar{W}(r_{ij},\hat{d}_i,\hat{d}_j)}=\langle e^{-\beta W(r_{ij},\hat{r}_{ij},\hat{d}_i,\hat{d}_j)} \rangle_{\hat{r}_{ij}}$, where $\langle \ \rangle_{\hat{r}_{ij}}$ denotes the ensemble average over the possible configurations of $\hat{r}_{ij}$, which, in the case of a liquid, coincides with an average over a uniform distribution on the solid angle. The bare potential $W(r_{ij},\hat{r}_{ij},\hat{d}_i,\hat{d}_j)$ can be the dipolar interaction itself or a renormalization thereof that effectively accounts for many-body interactions. In either case, the screened dipolar interaction $\bar{W}$ is well defined, only if the associated free energy
\begin{eqnarray}
	\bar{F}=-\frac{1}{\beta} \ln Z =\nonumber \\  -\frac{1}{\beta} \ln \langle e^{-\beta \sum_{i<j} [v_0(r_{ij})+\bar{W}(r_{ij},\hat{d}_i,\hat{d}_j)]} \rangle_{\{\mathbf{r}_i,\hat{d}_i\}} \label{F_liq_scr}, 
\end{eqnarray}
coincides, up to an irrelevant additive constant, with the free energy in Eq. \ref{F_liq}. The second issue raised in this study is the following:
\begin{enumerate}[resume*=ferro]
	\item  Can positional disorder in the liquid screen the dipolar interaction? More specifically, can annealed averaging over $\hat{r}_{ij}$ generate an isotropic, short-ranged dipolar potential that fully characterizes the free energy of the liquid?
\end{enumerate}
It is worth emphasizing that, in order to answer these questions, for a liquid, where the one-particle density is independent of $\boldsymbol{r}$, the free energy cannot be computed within the mean-field approximation, since within this framework the only nonvanishing contribution is the surface term. The question becomes thus ill posed, since no isotropic potential can reproduce a purely shape-dependent contribution. A first indication that such an effective potential may exist, possibly inducing a ferroelectric phase transition, is provided by the observation that both dipolar and Heisenberg fluids with purely short-range spin–spin interactions exhibit orientationally ordered phases and display similar phase diagrams \cite{Cattes, Weis1, Nijmeijer, Mryglod, Tavares}. 
A ferroelectric phase transition in a liquid implies dipolar ordering without crystallization, a condition expected to hold when the dipolar interaction is a sufficiently weak perturbation of an otherwise isotropic reference potential. The screened dipolar interaction introduced above can then be interpreted as arising from the screening of the perturbing dipolar potential by the isotropic reference potential, in analogy with screening mechanisms such as the random-phase approximation (RPA) for fluids \cite{Hansen}, Debye–Hückel screening in ionic solutions \cite{Hansen}, and optimized cluster expansion of the free energy \cite{Hansen,Chandler0}.

Addressing the questions raised above calls for a theoretical framework beyond the mean-field approximation. The role of dipolar pair correlations in the dielectric behavior of polar liquids was first recognized in the seminal study by Kirkwood \cite{Kirkwood}. Following Ref. \cite{Kirkwood}, the dipolar pair correlation function is given by the following formally exact expression (in the absence of ferroelectric order):
\begin{eqnarray}
	\langle \hat{d}_i \cdot \hat{d}_j\rangle=\frac{\int d \hat{d}_id \hat{d}_j  d \mathbf{r}_{ij}  \ \hat{d}_i \cdot \hat{d}_j \ e^{-\beta V(\mathbf{r}_{ij},\hat{d}_i,\hat{d}_j)}}{\int d \hat{d}_i \hat{d}_j d \mathbf{r}_{ij} e^{-\beta V(\mathbf{r}_{ij},\hat{d}_i,\hat{d}_j)}}. \label{W}
\end{eqnarray}
$V(\mathbf{r}_{ij},\hat{d}_i,\hat{d}_j )$ is the two-body potential of mean force between a pair of molecules with dipole orientations $\hat{d}_{i(j)}$ and center-of-mass distance $\mathbf{r}_{ij}$. It can be obtained by marginalizing the full configurational Boltzmann weight $e^{-\beta V_N(\{\mathbf{r}_{k}, \hat{d}_k\})}$ over all degrees of freedom other than $\hat{d}_{i(j)}$ and $\mathbf{r}_{ij}$. $V_N$ is the total potential energy of a configuration $\{\mathbf{r}_{k}, \hat{d}_k\}$. In the case of pairwise additive interactions described by a two-body potential $v(\mathbf{r}_{ij}, \hat{d}_i, \hat{d}_j)$, $V_N=\sum_{i<j}v(\mathbf{r}_{ij}, \hat{d}_i, \hat{d}_j)$. By its very definition, $V(\mathbf{r}_{ij},\hat{d}_i,\hat{d}_j)$ coincides, up to an additive constant, with $-\frac{1}{\beta}\log g^{(2)}\big(\mathbf{r}_{ij},\hat{d}_i,\hat{d}_j)$, where $g^{(2)}(\mathbf{r}_{ij},\hat{d}_i,\hat{d}_j)$ is the pair correlation function  appearing in Eq. \ref{F_DFT}. In the present case	$v(\mathbf r_{ij},\hat d_i,\hat d_j)=v_0(r_{ij})+w_p(\mathbf r_{ij},\hat d_i,\hat d_j)$. In schemes where many-body effects are reduced to effective pair interactions, the form $V(\mathbf r_{ij},\hat d_i,\hat d_j)=V_0(r_{ij})+W(\mathbf r_{ij},\hat d_i,\hat d_j)$ is typically retained, with $V_0$ a reference potential and $W$ a perturbative contribution.
In full generality, Eq. \ref{W} takes the form
\begin{eqnarray}
	\langle \hat{d}_i \cdot \hat{d}_j\rangle=\nonumber \\ \frac{\int d \hat{d}_id \hat{d}_j  d \mathbf{r}_{ij}  \ \hat{d}_i \cdot \hat{d}_j \ \tilde{\rho}(\mathbf{r}_i,\hat{d}_i)e^{-\beta V(\mathbf{r}_{ij},\hat{d}_i,\hat{d}_j)}\tilde{\rho}(\mathbf{r}_j,\hat{d}_j)}{\int d \hat{d}_i \hat{d}_j d \mathbf{r}_{ij} \tilde{\rho}(\mathbf{r}_i,\hat{d}_i)e^{-\beta V(r,\hat{r}_{ij},\hat{d}_i,\hat{d}_j)}\tilde{\rho}(\mathbf{r}_j,\hat{d}_j)}. \label{W1}
\end{eqnarray}	
The relevance of this latter quantity for the macroscopic polarization is readily appreciated by considering the square of $\bar{\mathbf{p}}$ in Eq. \ref{P}. 
From Eqs. \ref{W} and \ref{W1} the central role of the effective potential $V$ clearly emerges: it sets the hindered rotation of the dipoles, which is directly encoded in the pair correlation function $\langle \hat{d}_i \cdot \hat{d}_j \rangle$  \cite{Kirkwood}. A positive sign of this latter signals a tendency of dipole pair to align. This is readily illustrated by considering a Heisenberg fluid, i.e., a system of particles carrying vector spins, suppose they are the $\hat{d}_{i(j)}$ in Eq. \ref{W}, interacting via a short-ranged, isotropic ferromagnetic exchange potential, $V(\mathbf{r}_{ij},\hat{d}_i,\hat{d}_j)=J(r_{ij})\hat{d}_i \cdot \hat{d}_j$, with $J(r)>0$.
In this case, $\langle \hat{d}_i \cdot \hat{d}_j \rangle>0$ as a direct consequence of the fact that $V$ is minimized by parallel orientations of $\hat{d}_i$ and $\hat{d}_j$, a configuration that therefore has a larger weight than oppositely aligned orientations in the integral in Eq. \ref{W}. Any effective potential possessing this property, when acting on dipolar degrees of freedom, may be termed ferroelectric-like. These observations suggest that the behavior of $\langle \hat{d}_i \cdot \hat{d}_j \rangle$ in the paraelectric phase provides a criterion to assess the possibility of a ferroelectric phase transition.
An explicit evaluation of $W$ is nontrivial. In Ref. \cite{Kirkwood} it is approximate by the potential of average torque, $W_0$, acting between a pair of nearest-neighbor dipoles. Restricting the description to nearest neighbors, a notion of local structure is implicitly introduced. Consistently, in Ref. \cite{Kirkwood} $\langle \hat{d}_i \cdot \hat{d}_j \rangle$ depends on the average coordination number. When addressing the case of water, therefore, the focus is on tetrahedral coordination. No explicit microscopic characterization of $W_0$ is provided, and the origin of hindered rotation is ascribed to a combination of dipole–dipole electrostatic interactions and other intermolecular forces, whose relevance depends on the specific system. This approach cannot account for differences in the behavior of liquid and solid phases with similar local structure. 
In the case of water, for example, it cannot account for why the low-density liquid phase may display ferroelectric order, whereas hexagonal ice, despite a closely related local structure, does not. What distinguishes a liquid from a solid is exactly the presence, in the former, of annealed positional disorder. This brings us to the third, and central, question of this manuscript: 
\begin{enumerate}[resume*=ferro]
	\item  Can the screened dipolar potential $\bar{W}$, obtained from the annealed average over $\hat r_{ij}$ of the dipolar contribution $W$ to the effective two-body interaction, induce hindered dipole rotation leading to a ferroelectric phase transition in the liquid?
\end{enumerate}
If so, this would be a significant result: the driving mechanism for a bulk ferroelectric transition in dipolar liquids would then arise directly from the liquid character of the positional degrees of freedom, without invoking specific local structures or additional short-range interactions. It is worth emphasizing that the annealed averaging amounts to a marginalization of Eq. \ref{W} over $\hat{r}_{ij}$, and therefore does not represent a local property. This outcome would point to a distinct perspective on the ferroelectric phase transition in dipolar liquids, with potentially far-reaching implications.
For example, provided that water can be modeled as a dipolar liquid, the occurrence of a ferroelectric phase transition accompanying the high–to–low-density liquid phase transition in supercooled water \cite{Izzo} could be directly traced back to the liquid nature of the system itself, with the driving mechanism for ferroelectric order already present in the paraelectric high-density phase. By contrast, within a Kirkwood-like framework in which local structure governs hindered rotation, ferroelectric order could arise as a consequence of the specific local ordering in the low-density phase. In this view, ferroelectricity would follow the high– to low-density phase transition rather than drive it.
A second consequence is that frustration of dipolar orientations associated with specific lattice arrangements in solids, such as in certain ice phases, would be absent in the liquid, characterized by annealed positional disorder.

The rest of the paper is organized as follows. In Sec. \ref{cap2}, a dipolar liquid within a generalized Onsager reaction-field framework \cite{Onsager} is introduced, in which the shape-dependent surface contribution to the free energy vanishes. This enables an assessment of whether a genuine ferroelectric phase transition can occur, as posed in question (i).
As noted above, the marginalization of the full configurational Boltzmann weight $e^{-\beta V_N(\{\textbf{r}_k,\hat{d}_k\})}$ to derive $V(\mathbf{r}_{ij},\hat{d}_i,\hat{d}_j)$, or equivalently $g^{(2)}(\mathbf{r}_{ij},\hat{d}_i,\hat{d}_j)$, and then $W(\mathbf{r}_{ij},\hat{d}_i,\hat{d}_j)$ , is a nontrivial task.
A particularly transparent limit is obtained  when many-body correlations are negligible, as in the case where the virial expansion can be truncated at second order. In this case, $V(\mathbf{r}_{ij},\hat{d}_i,\hat{d}_j) = v(\mathbf{r}_{ij},\hat{d}_i,\hat{d}_j)$ and $W(\mathbf{r}_{ij},\hat{d}_i,\hat{d}_j) = w_p(\mathbf{r}_{ij},\hat{d}_i,\hat{d}_j)$. Since this truncation becomes exact in the limit $d \rightarrow \infty$, under suitable conditions on the thermodynamic state of the system and interaction parameters, it is natural to examine this case first. This is addressed in Sec. \ref{cap3}. The virial expansion of dipolar liquids in the limit $d \to \infty$, its reformulation within DFT, the conditions under which the truncation of the virial series becomes exact, and the scaling behaviour of the free energy in this limit are discussed in Sec. \ref{par3_1}. In the limit $d \to \infty$, the link between annealed averaging over $\hat{r}_{ij}$ and the emergence of ferroelectricity in dipolar liquids becomes particularly clear, as discussed in Sec. \ref{par3_2}. Issues (i)–(iii) can be addressed exactly in the limit $d \to \infty$, as shown in Sec. \ref{par3_3}. The analysis of the character of the screened dipolar potential, resulting from the annealed average of the bare dipolar interaction over $\hat{r}_{ij}$, its interaction range, and its role in hindering dipole rotation and shaping both dipolar and positional pair correlation functions, is presented in Sec. \ref{par3_4}. The generalization of the results obtained in the limit $d \to \infty$ to finite $d$ is addressed in Sec. \ref{cap4}.
In three-dimensional systems, truncation of the virial expansion at second order generally provides a poor approximation. It is therefore desirable to derive a bare effective two-body potential interaction between dipoles incorporating many-body effects, analogous to $V(\mathbf{r}_{ij},\hat{d}_i,\hat{d}_j)$ in Eq. \ref{W}, from which an effective free-energy expression with a second-order virial truncation can be constructed. The annealed averaging can then be performed on $W(\mathbf{r}_{ij},\hat{d}_i,\hat{d}_j)$, rather than on the bare dipolar potential. Sec. \ref{par4_2} introduces the so-called optimized cluster expansion for classical fluids \cite{Hansen, Chandler0}, which provides an approximate expression for $W$. Sec. \ref{par4_3} analyzes the properties of its annealed average over $\hat{r}_{ij}$. Sec. \ref{par4_4} derives a simplified three-dimensional expression for the screened dipolar interaction annealed-averaged over $\hat r_{ij}$, whose mean-field free energy qualitatively reproduces the tendency toward ferroelectric ordering obtained using the optimized cluster expansion. 
Although widely applicable \cite{Chandler0}, the optimized cluster expansion is an approximation. Results derived within this framework are therefore formally exact but valid only insofar as it provides a reliable description.
A simpler route would be to generalize the $d \rightarrow \infty$ results to finite $d$ in systems where the virial expansion truncates at second order and correlations beyond pairwise are negligible. In the Supplemental Material it is indeed shown that in this case the corresponding $\mathcal{F}$ is minimized by a ferroelectric state. However, the use of the optimized cluster expansion provides a broader scope for the present results, showing that they remain valid even when many-body correlations are incorporated into an effective pair potential. Conclusions are drawn in Sec. \ref{cap_concl}.

\section{Dipolar Interaction: The Intrinsic Bulk Contribution} \label{cap2}
The existence of a genuine bulk ferroelectric phase transition in dipolar liquids can be assessed only under conditions where the shape-dependent surface term is suppressed. A possible route is to consider the reaction-field construction, originally introduced by Onsager \cite{Onsager}. A generalization \cite{Leeuv, Neumann, Wei, Stenqvist, Bell} to more than one dipole inside the cavity is adopted here, following the approach of Kirkwood \cite{Kirkwood}. A spherical cavity of radius $R_c$ is carved out of the dipolar liquid, containing a certain number of dipoles. The liquid outside the cavity is modeled as a homogeneous and isotropic dielectric continuum with permittivity $\epsilon$, and it is assumed to respond linearly to the polarization inside the cavity. This construction induces an additional two-body effective interaction between dipoles inside the cavity, encoding many-body effects due to the mean-field response of the surrounding dielectric medium.
The spherical geometry of the cavity enforces an isotropic response, free from contributions associated with macroscopic surface anisotropies. In the limit $R_c \rightarrow \infty$, any macroscopic polarization must therefore originate from the bulk.
The dipolar liquid is governed by the potential in Eq. \ref{v_lambda}, with $\lambda = 1$. The dipolar interaction inside the cavity is described by the following generalized expression, incorporating the Onsager construction:
\begin{widetext}
\begin{multline}
	w_p(r, \hat{r}_{ij},\hat{d}_i, \hat{d}_j; R_c, \epsilon)=-p^2 \Big[\Big( \frac{l}{r} \Big)^{d} [d(\hat{d}_i \cdot \hat{r}_{ij})(\hat{d}_j \cdot \hat{r}_{ij})-\hat{d}_i\cdot \hat{d}_j]+\frac{f_d(\epsilon)}{R_c^d}\hat{d}_i\cdot \hat{d}_j \Big] \theta(r - l) \theta(R_c-r), \label{dippotddim}
\end{multline} 
\end{widetext}
where $f_d(\epsilon)>0$ is a bounded function of $\epsilon$, $\forall d$.
The last term in square brackets in Eq. \ref{dippotddim} describes the contribution to the dipolar interaction arising from the surrounding medium.
$\theta(x)$ is the Heaviside step function and $l$ is the hard-sphere diameter when $v_0$ is a hard-sphere potential, or the effective interaction range when $v_0$ is a Lenard-Jones potential. The factor $\theta(r - l)$ ensures that the perturbative potential vanishes within the core region, as discussed in Sec. \ref{intro}. 
The scaling factor $l^{d}$ sets the characteristic length scale associated with $w_p$, $L=l$.
To leading order in the limit $d \rightarrow \infty$, $w_p(r)$ is nonzero only within a narrow region around $r=L$ \cite{Parisi}, as can be readily inferred from its expression in Eq. \ref{dippotddim}.
Hence, if $L<l$, $w_p$ becomes ineffective as $d \rightarrow \infty$. If $L>l$, the leading large-$d$ behavior is the same as for $L=l$. Fixing $L=l$ is then the more natural choice. The above construction assumes $R_c>l$. The effective pair potential in Eq. \ref{dippotddim} satisfies both the stability and the temperedness conditions. 
Three remarks are in order regarding the contribution of the mean-field response of the surrounding medium to the dipolar pair interaction inside the cavity in Eq. \ref{dippotddim}, $-p^2 \frac{f_d(\epsilon)}{R_c^d} \hat{d}_i \cdot \hat{d}_j$.
(i) This term favors the alignment of two dipoles. This ferroelectric-like character, is a direct consequence of the assumption of a linear and isotropic response of the external medium, which determines the functional form of this contribution. 
(ii)
In the liquid phase, ergodicity of the positional degrees of freedom implies an isotropic distribution of the intermolecular direction $\hat r_{ij}$ over the solid angle. The annealed averaging over positional degrees of freedom is therefore performed accordingly. Assuming a spherical cavity, thereby modeling an isotropic response of the surrounding medium, is thus coherent with the annealed averaging of the medium inside the cavity. One can anticipate that this averaging likewise imparts a ferroelectric-like character to the dipolar interaction.
iii) In the limit $d \rightarrow \infty$ this contribution vanishes. This reflects the fact that many-body correlations disappear in this limit, as discussed in Sec. \ref{cap3}.

The bulk behavior is obtained in the limit $R_c \to \infty$. Since $\langle w^0_p\rangle_{\hat{r}_{ij}}=0$, as observed in Sec. \ref{intro}, it is straightforward to verify that
\begin{widetext}
\begin{eqnarray}
	\int r^{d-1} dr \ d\hat{r}_{ij} \ w_p(r, \hat{r}_{ij},\hat{d}_i, \hat{d}_j; R_c, \epsilon)=-p^2 \hat{d}_i \cdot \hat{d}_j   f_d(\epsilon)\frac{\Omega_d}{d}[1-\Big(\frac{l}{R_c}\Big)^d] . \label{wpint0}
\end{eqnarray}
\end{widetext}
In the limit $d \rightarrow \infty$, the right-hand side of Eq. \ref{wpint0} vanishes.
For any function $g$ of $w_p(r, \hat{r}_{ij},\hat{d}_i, \hat{d}_j; R_c, \epsilon)$ admitting a Taylor expansion around $w_p=0$, the following relation then holds: 
\begin{widetext}
\begin{multline}
	\lim_{R_c \rightarrow \infty}\int_0^{\infty} r^{d-1} dr \ \int_{\Omega_d} d\hat{r}_{ij} \   g\big(w_p(r, \hat{r}_{ij},\hat{d}_i, \hat{d}_j; R_c, \epsilon)\big)=   -C_{g}^d p^2 \hat{d}_i \cdot \hat{d}_j   f_d(\epsilon)\frac{\Omega_d}{d}+\Omega_d \int_0^{\infty} r^{d-1} d r \ \big\langle g\big(w_p(r, \hat{r}_{ij},\hat{d}_i, \hat{d}_j)\big) \big\rangle_{\hat{r}_{ij}}, \label{averagerij}
\end{multline} 
\end{widetext}
Eq. \eqref{averagerij} follows by owing to Eq. \ref{wpint0} and expanding $g$ in a Taylor series in $w_p$. Once using the mean reaction-field construction, each term of the expansion yields an absolutely convergent integral, allowing the exchange of the limit and the integration. The constant $C_g^d$ stems from the linear term in the Taylor expansion. As $d \to \infty$, the first term in Eq. \ref{averagerij} vanishes.

\section{Bulk Ferroelectricity of Dipolar Liquids in Large Dimensions} \label{cap3}
\subsection{Virial Expansion and Density-Functional Theory} \label{par3_1}

The virial expansion is a high-temperature expansion which can also be interpreted as a large $d$ expansion \cite{Parisi}. 
In the limit $d\to\infty$, the excess free energy of a hard-sphere
liquid with respect to the ideal gas reduces, under suitable conditions, to the second virial
term, corresponding to direct two-particle interaction \cite{Parisi}.
In this limit, contributions to the free energy arising from correlations beyond the pair level vanish. This result is valid as long as the packing fraction of the liquid satisfies a suitable upper-bound condition \cite{Parisi}, ensuring that the second virial coefficient remains finite. If this condition is violated, the virial expansion diverges and higher-order virial coefficients, associated with correlations beyond the pair level, become dominant.
The dipolar interaction in Eq. \ref{dippotddim}, although anisotropic in $\mathbf{r}$-space, is still rotationally invariant in the extended configuration space including both translational and dipolar degrees of freedom. The arguments of Ref. \cite{Parisi} establishing the exactness of the second-order truncation of the virial series can therefore be extended to the potential in Eq. \ref{dippotddim} \cite{Yoshino}.
Including dipolar degrees of freedom makes the radius of convergence of the virial expansion dependent on parameters associated with these additional degrees of freedom. The conditions of validity of the second-order truncation in the limit $d \to \infty$ therefore translate into corresponding constraints on such parameters, as discussed below.

The central object of the virial expansion is the Mayer function. For the potential $v$ in Eq. \ref{v_lambda} with $\lambda=1$, and the dipolar potential in Eq. \ref{dippotddim}, it reads
\begin{eqnarray}
	f(r, \hat{r}_{ij},\hat{d}_i, \hat{d}_j;R_c, \epsilon)=e^{-\beta [v_0(r)+w_p(r, \hat{r}_{ij},\hat{d}_i, \hat{d}_j;R_c,\epsilon)]}-1, \nonumber \\ \label{Mayer}
\end{eqnarray}
The large-$d$ limit of the corresponding free energy $F[\tilde{\rho}]$ is obtained by truncating the virial expansion at second order and taking the limit $R_c\to\infty$. One finds
\begin{widetext}
\begin{eqnarray}
	-\beta F[\tilde{\rho}]=-\beta F^{id}[\tilde{\rho}]+\frac{N\rho}{2}\Omega_d \int r^{d-1} dr d  \hat{d}_i d  \hat{d}_j \ \zeta(\hat{d}_i) \langle f(r, \hat{r}_{ij},\hat{d}_i, \hat{d}_j) \rangle_{\hat{r}_{ij}} \zeta(\hat{d}_j), \label{Fsecondvirial}
\end{eqnarray} 
\end{widetext}
where $F^{id}[\tilde{\rho}]$ is the ideal-gas free energy. Eq. \ref{Fsecondvirial} follows from Eq. \ref{averagerij} by choosing $g=f$. Within the replicated liquid theory framework \cite{Parisi}, Eq. \eqref{Fsecondvirial} refers to the single-replica (liquid), and the average over $\hat{r}_{ij}$ corresponds to an annealed average. At second virial order, the excess free energy is a linear functional of $f$. This makes it straightforward to introduce an effective two-body screened dipolar potential obtained after carrying out the annealed average over $\hat{r}_{ij}$ in Eq. \ref{Fsecondvirial}, as discussed in Secs. \ref{par3_2}-\ref{par3_3}. 
Eq. \ref{Fsecondvirial} already provides a representation of the free energy in terms of the one-particle density field and therefore naturally falls within the DFT framework. It can be recast in the standard DFT expression, separating contribution to the free energy arising from a reference system and a perturbative potential. Decomposing the Mayer function in Eq. \ref{Mayer} as
\begin{eqnarray}
	f(r, \hat{r}_{ij},\hat{d}_i, \hat{d}_j;R_c,\epsilon) =\nonumber \\ f_{v_0}(r)+e^{-\beta v_0(r)}f_{w_p}(r, \hat{r}_{ij},\hat{d}_i, \hat{d}_j;R_c,\epsilon), \label{Mayer_vowp}
\end{eqnarray}
with $f_{v_0}= e^{-\beta v_0}-1$ and $f_{w_p}= e^{-\beta w_p}-1$, Eq. \ref{Fsecondvirial} reduces to
\begin{eqnarray}
	F[\tilde{\rho}]=F_{v_0}[\tilde{\rho}]+\mathcal{F}[\tilde{\rho}]. \label{F_DFT0}
\end{eqnarray}
$F_{v_0}[\tilde{\rho}]$ is the free energy of the reference system,  describing particles interacting through $v_0$ and carrying non-interacting dipoles, while $\mathcal{F}$ is the excess free energy associated to $w_p$,
\begin{align}
	\mathcal{F}[\tilde{\rho}]
	&=
	-\frac{N\rho}{2\beta}\Omega_d
	\int r^{d-1}dr\, e^{-\beta v_0(r)}
	d\hat d_i d\hat d_j\,
	\zeta(\hat d_i)
	\nonumber\\
	&\quad\times
	\langle
	f_{w_p}(r,\hat r_{ij},\hat d_i,\hat d_j)
	\rangle_{\hat r_{ij}}
	\zeta(\hat d_j).
	\label{Fexdinf}
\end{align}
Eq. \ref{Fexdinf} provides the exact DFT expression for the excess free energy in the limit $d \rightarrow \infty$.

In the limit $d\to\infty$, the entropic cost for breaking orientational isotropy scales as $O(d)$, as shown in point (ii) of Appendix \ref{appendixI}. Therefore, in order for the dipolar interaction to produce a non-trivial contribution to the excess free energy and compete with the entropic term, its magnitude must scale accordingly. This is achieved by assuming
\begin{equation}
	p^2=d \bar{p}^2 \label{prescale}
\end{equation}
With this scaling the dipolar interaction remains finite only within a thin boundary layer around $r=l$, of thickness $O(l\log d/d)$, as shown in point (iii) of Appendix \ref{appendixI}, which is conveniently parametrized as,
\begin{equation}
	r = l\left(1+\frac{\log d+h}{d}\right), \ \ \ \ h=O(1).
	\label{double_scaling}
\end{equation}
In this region indeed it is
\begin{equation}
	\beta p^2\left(\frac{l}{r}\right)^d
	\;\xrightarrow[d\to\infty]{}\;
	\beta\bar p^{\,2}e^{-h}.
	\label{finite_coupling}
\end{equation}
However, as shown in Appendix \ref{appendixI}-(iii), under the scaling in Eq. \ref{prescale} the dipolar interaction can become unbounded in the limit $d\to\infty$ whenever $h<0$ with sufficiently large $|h|$, corresponding to the region $l<r<l\left(1+\frac{\log d}{d}\right)$, possibly inducing instability in the system. As a consequence, the large-$d$ limit is well defined only outside this region. This can be implemented by introducing an effective core at $l_{\mathrm{eff}} = l\left(1+\frac{\log d}{d}\right)$, which replaces $l$ in the Heaviside function in Eq. \ref{dippotddim}, so that the condition $r>l_{\mathrm{eff}}$ becomes equivalent to $h>0$, following Eq. \ref{double_scaling}. This directly yields the Heaviside function $\theta(h)$ in Eq. \ref{fbarexp}. Finally, in order to obtain a non-trivial contribution from the isotropic potential $v_0(r)$, its amplitude is assumed to scale with $d$, as for
the dipolar interaction, so that
under the radial scaling in Eq. \ref{double_scaling} one obtains $v_0(r)\rightarrow  v_0(h)$, with $v_0(h)$ finite for $d\to\infty$. As detailed in Appendix \ref{appendixI}-(iii), the excess free-energy in the limit $d \rightarrow \infty$ thus reduces to
\begin{eqnarray}
	\mathcal{F} [\tilde{\rho}]
	=
	N\rho B_2^{HS}
	\int d \hat{d}_i\, d \hat{d}_j\;
	\zeta(\hat{d}_i)\, I(\hat{d}_i \cdot \hat{d}_j)\,\zeta(\hat{d}_j),
	\label{FIh}
\end{eqnarray}
where $B_2^{HS}=\Omega_d l^d/(2d)$ is the hard-sphere second virial coefficient, and the reduced density $\rho B_2^{HS}$ is kept finite in the limit $d\to\infty$.
The kernel $I(\hat{d}_i \cdot \hat{d}_j)$ is given by
\begin{eqnarray}
	I(\hat{d}_i \cdot \hat{d}_j)
	=
	-\frac{d}{\beta}\int_{-\infty}^{\infty}dh \;
	e^h e^{-\beta {v}_0(h)}\,
	\bar{f}_{w_p}(\hat{d}_i \cdot \hat{d}_j,h),
	\label{Ih}
	\\
	\bar{f}_{w_p}(\hat{d}_i \cdot \hat{d}_j,h)
	=
	\lim_{d \rightarrow \infty}
	\langle f_{w_p}(r, \hat{r}_{ij},\hat{d}_i, \hat{d}_j) \rangle_{\hat{r}_{ij}}.
	\label{fbar}
\end{eqnarray}
Making Eq. \ref{fbar} explicit, see Appendix \ref{appendixI}-(iii),
\begin{widetext}
\begin{eqnarray}
	\bar{f}_{w_p} (\hat{d}_i \cdot \hat{d}_j,h)=\lim_{d \rightarrow \infty} \frac{1}{\Omega_d} \int d\hat{r}_{ij}  e^{-\beta \bar{p}^2[d(\hat{d}_i \cdot \hat{r}_{ij})(\hat{d}_j \cdot \hat{r}_{ij})-\hat{d}_i\cdot \hat{d}_j]\theta(h)e^{- h}}-1.  \label{fbarexp}
\end{eqnarray}
\end{widetext}
Notice that, after averaging $f_{w_p}(r,\hat r_{ij},\hat d_i,\hat d_j)$ over the uniform solid-angle distribution of $\hat r_{ij}$, the resulting function $\bar f(r,\hat d_i,\hat d_j)$ depends only on the scalar product $\hat d_i\cdot\hat d_j$, as explicitly shown in Sec. \ref{par3_3}.
If the Mayer function $\bar f_{w_p}(\hat d_i \cdot \hat d_j,h)$ remains uniformly bounded, and $v_0$ is tempered, as in the present case, the virial series terms of order $n \ge 3$ are subleading in the limit $d \to \infty$ at fixed reduced density $\rho B_2^{HS}=O(1)$ \cite{Parisi}. The condition of uniformly boundedness of $\bar f_{w_p}(\hat d_i \cdot \hat d_j,h)$ imposes a threshold on $\beta \bar{p}^2$, as shown in Sec. \ref{par3_3} and Appendix \ref{appendixI}-(v) to which the validity of the second-order truncation of the virial series is therefore restricted. Under this condition, the standard large-$d$ exactness proof for hard spheres \cite{Parisi} extends straightforwardly to the potential $v_0+w_p$. As can be inferred from Eqs. \ref{FIh}-\ref{fbarexp}, the excess free energy in the limit $d \rightarrow \infty$ retains a specific dependence on the dipole-orientation probability distribution, entirely encoded in $\bar{f}_{w_p} (\hat{d}_i \cdot \hat{d}_j,h)$. Furthermore, the boundedness of $\bar f(r,\hat d_i \cdot \hat d_j)$, implies that $\mathcal{F}=O(d)$ in the limit $d\to\infty$.

Owing to the general identity
\begin{eqnarray}
	\frac{\delta F[\tilde{\rho}]}{\delta v(\textbf{r}_i, \textbf{r}_j, \hat{d}_i, \hat{d}_j) }=-\frac{1}{2} \rho^{(2)}(\textbf{r}_i,\textbf{r}_j,\hat{d}_i,\hat{d}_j), \label{hr_der}
\end{eqnarray} 
it follows immediately that, upon truncating the virial expansion at second order and using the one-particle density field in Eq. \ref{rho_om}, the pair correlation function reads
\begin{equation}
	g^{(2)}(r, \hat{r}_{ij},\hat{d}_i,\hat{d}_j)= e^{-\beta [v_0(r)+w_p(r, \hat{r}_{ij},\hat{d}_i, \hat{d}_j)]}. \label{pair_lowrho}
\end{equation}  
The same ansatz in Eq. \ref{pair_lowrho} has been used in Refs. \cite{Teixeira, Frodl, Groh, Szalai, Gramzow} for dipolar liquids and in Refs. \cite{Tavares, Weis, Lomba, Schoen, Cattes} for classical Heisenberg fluids, and referred to as so-called modified mean-field approximation. In these studies, its use is justified in the low-density limit.
In Ref. \cite{Groh}, the use of the ansatz in Eq. \ref{pair_lowrho} is shown to lead to a ferroelectric phase transition in dipolar liquids. However, the analysis is based on the bare dipolar interaction, without including an Onsager-like reaction-field construction, and the pair correlation function in Eq. \ref{pair_lowrho} is expanded for small molecular dipole magnitude $p$, retaining only the leading terms. As a consequence, the dominant contribution to the free energy driving the ferroelectric phase transition originates from the surface term. The study therefore does not clarify whether the ferroelectric phase transition is a genuine bulk phenomenon.
Furthermore, while the ansatz in Eq. \ref{pair_lowrho} becomes exact as $d \to \infty$, it remains approximate in three dimensions, as observed in Sec. \ref{intro}.

\subsection{Annealed Positional Disorder and The Emergence of Ferroelectricity} \label{par3_2}
Eq. \ref{Fexdinf} shows that, in the limit $d\to\infty$, the free energy of the dipolar liquid is equivalent to that obtained from the effective pair Boltzmann factor 
\begin{equation} \langle e^{-\beta[v_0(r_{ij})+w_p(r_{ij},\hat r_{ij},\hat d_i,\hat d_j)]}\rangle_{\hat r_{ij}}. \label{effBF}
\end{equation} 
This follows directly from the exactness, in the large-$d$ limit, of the truncation of the virial expansion at second order, implying that only pair interactions contribute to the free energy. Otherwise, the free energy would also contain annealed averages involving products of Boltzmann factors associated with different particle pairs. To clarify the role played by the annealed average over $\hat r_{ij}$ in the emergence of ferroelectricity in dipolar liquids in the large-$d$ limit, it is useful to consider the expression of the liquid free energy in terms of the partition function, Eq. \ref{F_liq}.
Since, in the large-$d$ limit, the free energy of the dipolar liquid is entirely determined by the effective Boltzmann factor in Eq.~\ref{effBF}, the following relation holds:
\begin{align}
	Z&\equiv  \big\langle \prod_{i<j}e^{-\beta[v_0(r_{ij})+w_p(r_{ij},\hat{r}_{ij},\hat{d}_i,\hat{d}_j)]}\big\rangle_{\{\mathbf{r}_i,\hat{d}_i\}}
	\nonumber\\
	&\propto \big\langle\prod_{i<j}\langle e^{-\beta[v_0(r_{ij})+w_p(r_{ij},\hat{r}_{ij},\hat{d}_i,\hat{d}_j)]}\rangle_{\hat{r}_{ij}}\big\rangle_{\{\textbf{r}_{i},\hat{d}_i\}}, 
	\label{Z_an}
\end{align}
where the proportionality constant is physically irrelevant. In Eq. \ref{Z_an}, the annealed average in the right-hand side could, in principle, be extended from  $\hat r_{ij}$ to the full vectors $\mathbf r_{ij}$ or to the dipole
orientations $\hat d_i$, without affecting the formal validity of the
relation. However, the physical mechanism underlying the ferroelectric phase transition is encoded in the annealed average over $\hat{r}_{ij}$, which, through the functional form of $w_p$, controls the effective dipole–dipole coupling. 
Nevertheless, the observation above emphasizes that this average only represents an intermediate mathematical step introduced to isolate the physically relevant features of the problem. It is not an approximation and, in particular, it does not rely on any separation of time scales between translational and dipolar degrees of freedom. 

It is instructive to draw an analogy with the annealed counterpart of the Sherrington-Kirkpatrick spin-glass model \cite{SK}. Let a system of interacting vector spins $\hat d_i$ with
Hamiltonian
\begin{equation}
	H=
	\sum_{i<j}
	J_{ij}\,\hat d_i\cdot\hat d_j, \label{H_SK}
\end{equation}
where the couplings $J_{ij}$ are independently sampled from a
zero-mean Gaussian distribution.
The partition function of the annealed system can be written as
\begin{align}
	Z &\equiv \big\langle \prod_{i<j} e^{-\beta\, J_{ij}\,\hat d_i \cdot \hat d_j} \big\rangle_{\{J_{ij},\hat d_i\}}
	\nonumber\\
	&\propto 
	\big\langle \prod_{i<j} \langle e^{-\beta\, J_{ij}\,\hat d_i \cdot \hat d_j} \rangle_{J_{ij}} \big\rangle_{\{\hat d_i\}},
	\label{Z_anSK}
\end{align}
where $\langle \ \rangle_{\{J_{ij},\hat{d}_i\}}$ is the ensemble average over both coupling realizations and spin configurations, 
$\langle \ \rangle_{J_{ij}}$ is the average over the distribution of $J_{ij}$, 
and $\langle \ \rangle_{\{\hat{d}_i\}}$ is the ensemble average over spin configurations. In this case, Eq. \ref{Z_anSK} follows from the pairwise structure of the Hamiltonian and from the statistical independence of the couplings $J_{ij}$. The analogy with Eq. \ref{Z_an} is immediate, with $\hat r_{ij}$ playing a role formally analogous to the random couplings $J_{ij}$.
Interestingly, the annealed Sherrington-Kirkpatrick spin-glass model is known to exhibit a hidden Mattis phase with spin order \cite{Matsuda, Kasai}.

Eqs. \ref{Fexdinf} and \ref{Z_an} show that, in the limit $d \rightarrow \infty$, a screened dipolar potential $\bar{W}$ satisfying Eq. \ref{F_liq_scr} can be defined through the annealed average of the dipolar potential $w_p$ over $\hat{r}_{ij}$.
It reads as
\begin{eqnarray}
	\bar{W}_{d\to\infty}(r,\hat{d}_i,\hat{d}_j)=-\frac{1}{\beta}\lim_{d\to\infty}\log \langle e^{-\beta w_p(r,\hat{r}_{ij},\hat{d}_i,\hat{d}_j)}\rangle_{\hat{r}_{ij}} \label{Wbar_dinf}
\end{eqnarray}
To address the issues raised in Sec. \ref{intro} the remaining step is to determine whether the screened potential $\bar{W}$ in Eq. \ref{Wbar_dinf} is feroelectric-like, according to the definition introduced in Sec. \ref{intro}, and short-ranged.
Before proceeding to a fully quantitative analysis, useful qualitative insight can be obtained by considering the $d$-dimensional generalization of an octahedral quadrature scheme.
Within this approximation, 
\begin{eqnarray}
	e^{-\beta \bar W_{oq,d}(r,\hat d_i,\hat d_j)}
	=
	\frac{1}{2d}
	\sum_{\alpha=1}^{d}
	\sum_{\sigma=\pm 1}
	e^{-\beta w_p(r,\sigma \hat e_k,\hat d_i,\hat d_j)}, \label{Wbaroq}
\end{eqnarray}
where $\{\hat e_k\}_{\alpha=1}^{d}$ denotes the canonical basis
of $\mathbb R^d$. The sum runs over the $2d$ vertices of the
$d$-dimensional cross-polytope. Choosing one quadrature direction parallel to $\hat d_i$, Eq. \ref{Wbaroq} reduces to
\begin{widetext}
\begin{eqnarray}
	e^{-\beta \bar W_{oq,d}(r,\hat d_i,\hat d_j)}
	=
	\frac{1}{2d}
	\left[
	2e^{\beta(d-1)p^2\left(\frac lr\right)^d
		\hat d_i\cdot\hat d_j}
	+
	2(d-1)e^{-\beta p^2\left(\frac lr\right)^d
		\hat d_i\cdot\hat d_j}
	\right].
\end{eqnarray}
\end{widetext}
In the large-$d$ limit,
\begin{widetext}
\begin{eqnarray}
	\bar W_{oq,d\rightarrow \infty}(r,\hat d_i,\hat d_j)
	\sim
	-\frac{1}{\beta}
	\log
	\left[
	2e^{\beta(d-1)p^2\left(\frac lr\right)^d
		\hat d_i\cdot\hat d_j}
	+
	2(d-1)e^{-\beta p^2\left(\frac lr\right)^d
		\hat d_i\cdot\hat d_j}
	\right], \label{Wbaroqd}
\end{eqnarray}
\end{widetext}
where the additive term $-\log(2d)$ has been omitted, since it is orientation-independent. Minimizing the effective potential is equivalent to maximizing the quantity inside square brackets in Eq. \ref{Wbaroqd}. The first term is exponentially increasing with $\hat d_i\cdot\hat d_j$, whereas the second is exponentially decreasing. In the large-$d$ limit, the growth of the first term dominates, and the maximum is therefore attained for $\hat d_i\cdot\hat d_j=1$, implying that $\bar W_{oq}^{(d)}$ is minimized in the ferroelectric configuration.
In the limit $d \rightarrow \infty$ the minimum of $\bar{W}$ at $\hat{d}_i \cdot \hat{d}_j=1$ is accompanied by an unphysical divergence. This originates from the octahedral quadrature, which does not correctly reproduce the large-$d$ solid-angle measure and assigns finite weight to angular regions whose measure vanishes in the limit $d\to\infty$.  

The logarithm in Eq. \ref{Wbar_dinf} can be expanded as $\log(1+x)=x+O(x^2)$, with
$x=\langle e^{-\beta w_p} \rangle_{\hat r_{ij}}-1=-\beta\langle w_p\rangle_{\hat r_{ij}} +\frac{\beta^2}{2} \langle w_p^2\rangle_{\hat r_{ij}} +\cdots$, where $e^{-\beta w_p}=1-\beta w_p+\frac{\beta^2}{2}w_p^2+\cdots$ has been used. Since $\langle w_p\rangle_{\hat r_{ij}}=0$, as shown in Sec. \ref{intro}, the leading contribution is therefore proportional to $\langle w_p^2\rangle_{\hat r_{ij}}$.
This indeed anticipates that the screened potential $\bar{W}$ is shorter-ranged than the bare dipolar potential $w_p$.

\subsection{On the Onset of Ferroelectricity: An Exact Result} \label{par3_3}
As shown in Sec. \ref{par3_1} the quantity $\bar{f}_{w_p}(\hat{d}_i \cdot \hat{d}_j, h)$ defined in Eq. \ref{fbarexp} fully determines $\mathcal{F}$ in the limit $d \rightarrow \infty$. Once $\bar f_{w_p}(\hat d_i \cdot\hat d_j,h)$ is explicitly evaluated, the free energy is completely specified as a functional of the one-particle density field $\rho\,\zeta(\hat d)$. Minimization with respect to $\zeta(\hat{d})$ determines whether ferroelectric order can set in. This is the task addressed in the following. 

\begin{figure}[t]
	\centering
	\includegraphics[width=1\linewidth]{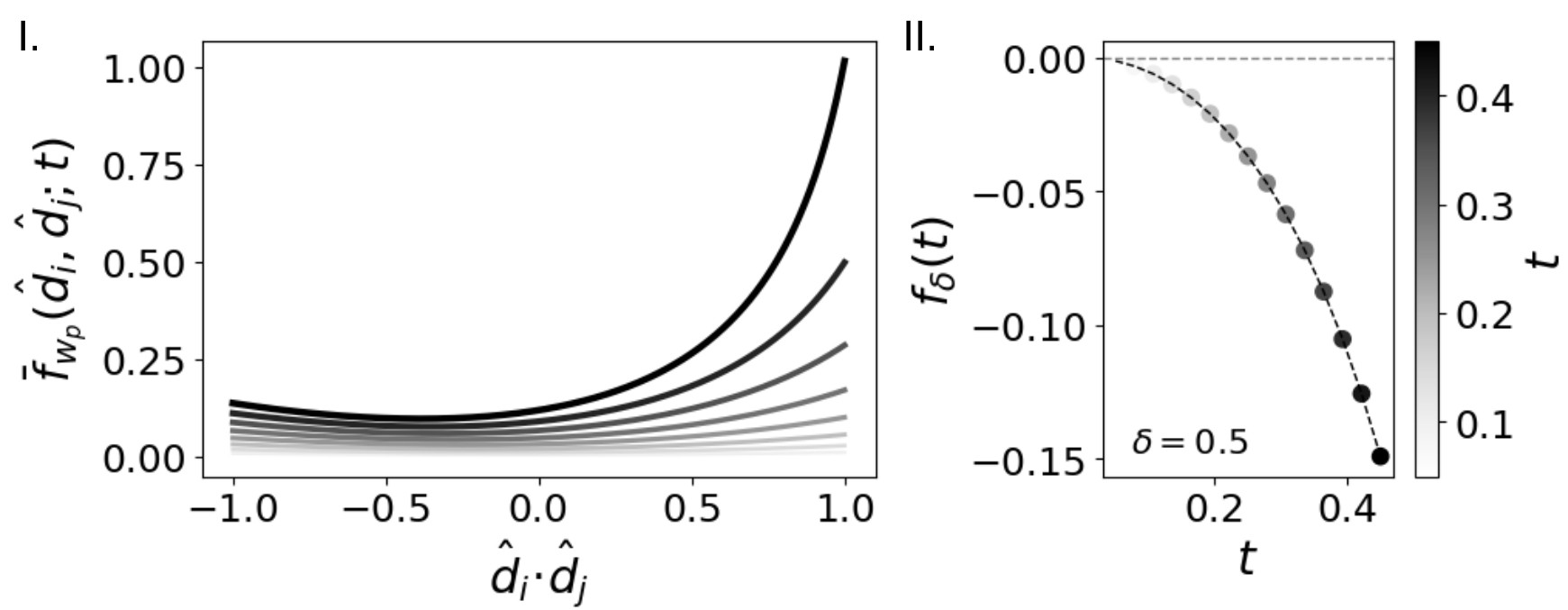}
	\caption{\textit{Panel I.} $\bar f_{w_p}(\hat d_i \cdot \hat d_j,t)$ as a function of $\hat{d}_i\cdot\hat{d}_j$ for different values of $t$. Increasing $t$ leads to progressively larger bias towards $\hat{d}_i\cdot\hat{d}_j=1$, indicating an enhanced ferroelectric-like character of $\bar{W}_{d \rightarrow \infty}$. At fixed $\beta$ and $h$, larger values of $t$ correspond to larger values of $p$. \textit{Panel II.} Corresponding values of $f_{\delta}$ as a function of $t$. $\forall t$, $f_{\delta}<0$.}
	\label{fig3}
\end{figure}

Changing variables from $\hat{r}_{ij}$ to $\mathbf{t}_{ij}=(\theta_i,\theta_j)=(\hat{d}_i \cdot \hat{r}_{ij},\hat{d}_j \cdot \hat{r}_{ij})$, one obtains
\begin{widetext}
\begin{eqnarray}
	\langle f_{w_p}(r, \hat{r}_{ij},\hat{d}_i, \hat{d}_j) \rangle_{\hat{r}_{ij}}= \frac{\Omega_{d-2}}{\Omega_{d}}\int_{\mathcal{D}} \frac{d \mathbf{t}_{ij}}{(\det \mathbf{G})^{\frac{1}{2}}}(1-\mathbf{t}_{ij}\mathbf{G}^{-1}\mathbf{t}_{ij}^T)^{\frac{d-4}{2}}e^{-\beta p^2\big( \frac{l}{r}\big)^{ d}[\hat{d}_i \cdot \hat{d}_j-d\theta_i \theta_j]}-1,  \label{fwp_annealed1}
\end{eqnarray} 
\end{widetext}
where $\mathcal{D} =\{ (\mathbf{t}_{ij}:\mathbf{t}_{ij}\mathbf{G}^{-1}\mathbf{t}_{ij}^T\leq 1 \}$, $\mathbf{t}_{ij}^T$ is the transpose of $\mathbf{t}_{ij}$, and $G$ is the $(2 \times 2)$ Gram matrix with entries $G_{ij}=\hat{d}_i \cdot \hat{d}_j$. Details are given in Appendix \ref{appendixI}-(v). Eq. \ref{fwp_annealed1} shows that $\langle f_{w_p}(r,\hat r_{ij},\hat d_i,\hat d_j)\rangle_{\hat r_{ij}}$ depends on $\hat d_i$ and $\hat d_j$ only through their scalar product, as anticipated in Sec. \ref{par3_1}.
To evaluate the limit $d \rightarrow \infty$ it is convenienet to introduce the rescaled variable $\tilde{\mathbf{t}}_{ij}=\sqrt{d}\mathbf{t}_{ij}$. As in Appendix \ref{appendixI}-(v), in the limit $d \rightarrow \infty$ one obtains
\begin{widetext}
\begin{eqnarray}
	\frac{\Omega_{d-2}}{\Omega_{d}}\frac{1}{(\det \mathbf{G})^{\frac{1}{2}}}(1-\mathbf{t}_{ij}\mathbf{G}^{-1}\mathbf{t}_{ij}^T)^{\frac{d-4}{2}} d \boldsymbol{t}_{ij} \ \underset{d\to\infty}{\longrightarrow} \   \ \frac{1}{2 \pi (\det \mathbf{G})^{\frac{1}{2}}} e^{-\frac{1}{2}\tilde{\mathbf{t}}_{ij}\mathbf{G}^{-1}\tilde{\mathbf{t}}_{ij}^T} d \tilde{\boldsymbol{t}}_{ij}, \label{transf}
\end{eqnarray}
\end{widetext}
with $\Omega_d=\frac{{2 \pi}^{\frac{d}{2}}}{\Gamma(\frac{d}{2})}$, and $\Gamma( \ )$ the Euler Gamma function. The probability distribution of the rescaled variables $\tilde{\mathbf t}_{ij}$, induced by the uniform measure on $\hat r_{ij}$, therefore converges for $d\to\infty$ to a centered bivariate Gaussian distribution with covariance matrix $\mathbf G$. 
Using the scaling variable $h$ for $r$, see Eq. \ref{fbarexp}, one finally obtains
\begin{widetext}
\begin{eqnarray}
	\bar{f}_{w_p}(\hat{d}_i \cdot \hat{d}_j,h)= \int_{-\infty}^{\infty} d \tilde{\mathbf{t}}_{ij}\frac{e^{-\frac{1}{2}\tilde{\mathbf{t}}_{ij}\mathbf{G}^{-1}\tilde{\mathbf{t}}_{ij}^T}}{2 \pi (\det \mathbf{G})^{\frac{1}{2}}}e^{-\beta \bar{p}^2[\hat{d}_i \cdot \hat{d}_j-\tilde{\theta}_i \tilde{\theta}_j]\theta(h)e^{-h}} -1.  \label{fbardinf}
\end{eqnarray} 
\end{widetext}
The integral in Eq. \ref{fbardinf} converges for all $h\ge 0$ provided $\beta \bar{p}^{2}<\tfrac{1}{2}$, thereby identifying the necessary condition for the second-order truncation of the virial series to be exact in the limit $d \to \infty$. 
For sufficiently large values of $\bar{p}$ higher-order orientational correlations among dipoles indeed emerge. Through the dipolar interaction, these correlations also induce effective correlations among particle positions, causing higher-order terms in the virial expansion to become dominant.
The integral in Eq. \ref{fbardinf} coincides with the moment-generating function $M_{\tilde{Z}}(t)$ of the random variable
\begin{equation}
	\tilde{Z}=[\tilde{\theta}_i \tilde{\theta}_j-\hat{d}_i \cdot \hat{d}_j], \label{Ztilde}
\end{equation}
evaluated with respect to the bivariate Gaussian distribution of $\mathbf{t}_{ij}$ defined above. Eq. \ref{fbardinf} can thus be written as
\begin{eqnarray}
	\bar{f}_{w_p}(\hat{d}_i \cdot \hat{d}_j,h)=\langle e^{t\tilde{Z}} \rangle_{\hat{r}_{ij}} -1=M_{\tilde{Z}}(t)-1,  \label{fbardinf_1}
\end{eqnarray}     
with $t=\beta \bar{p}^2 \theta(h)e^{-h}$ and therefore $t<\tfrac{1}{2}$ under the convergence
condition. As shown in Appendix \ref{appendixI}-(v),
\begin{equation}
	M_{\tilde{Z}}(t)=e^{-t\hat{d}_i \cdot \hat{d}_j-\frac{1}{2}\log[1-2 t\hat{d}_i \cdot \hat{d}_j -t^2(1-{\hat{d}_i \cdot \hat{d}_j}^2)]}. \label{MGF}
\end{equation}
Because $\langle w_p \rangle_{\hat{r}_{ij}}=0$, $\langle \tilde{Z} \rangle_{\hat{r}_{ij}}=0$, Jensen's inequality \cite{Jensen} then implies $M_{\tilde{Z}}(t)=\langle e^{t\tilde Z}\rangle_{\hat r_{ij}}\ge e^{t\langle \tilde Z\rangle_{\hat r_{ij}}}=1$, $\forall t$.
From Eqs. \ref{FIh}-\ref{fbarexp} and \ref{fbardinf_1}-\ref{MGF}, minimizing $\mathcal F$ amounts to maximizing $M_{\tilde Z}(t)$. A direct calculation from Eq. \ref{MGF} shows that $M_{\tilde Z}(t)$, as a function of $\hat d_i\cdot\hat d_j$, attains its maximum at $\hat d_i\cdot\hat d_j=1$.
This behavior is readily confirmed by inspection of Fig. \ref{fig3}, which displays $\bar{f}_{w_p}(\hat{d}_i \cdot\hat{d}_j,t)=M_{\tilde{Z}}(t)-1$ as a function of $\hat{d}_i \cdot \hat{d}_j$ for different values of $t$. In particular, for fixed $\beta$ and $h$, increasing values of $t$ correspond to increasing values of the dipole moment $\bar{p}$. $\mathcal{F}$ is therefore minimized by the single-particle dipolar orientational distribution $\zeta(\hat d)$ in which all dipoles align along a common direction. On the other hand,  $F_{v_0}[\tilde{\rho}]$  includes the orientational entropy of non-interacting dipoles, which is maximized by an isotropic orientational distribution. 
The competition between this entropic contribution and the excess free-energy $\mathcal{F}$, whose balance is tuned by the thermodynamic parameters of the liquid, determines the onset of the ferroelectric phase transition in the dipolar liquid \cite{Izzo}. 
To make the discussion more quantitative, the simplified ansatz
\begin{equation}
	\zeta(\hat d)= \frac{1}{Z_d(\boldsymbol{\delta})} e^{d\boldsymbol{\delta}\cdot \hat d}, \label{zeta}
\end{equation}
is considered, with $Z_d(\boldsymbol{\delta})=\int d\hat d \  e^{\boldsymbol{d\delta}\cdot \hat d}$. 
As shown in Appendix \ref{appendixI}-(i), in the limit \(d\to\infty\) this ansatz is able to discriminate between the paraelectric and ferroelectric phases.
In particular,
\begin{eqnarray}
	\boldsymbol{\bar{p}}=\bar{u}(\delta) \hat{\delta}, \nonumber \\
	\bar{u}(\delta)=\big(\sqrt{1+4\delta^2}-1\big)/(2\delta). \label{pdelta}
\end{eqnarray}
For small $\delta$, Eq. \ref{pdelta} yields $\boldsymbol{\bar p}\propto \boldsymbol{\delta}$, showing that $\boldsymbol{\delta}$ plays the role of the ferroelectric order parameter.
The entropy difference between the dipole-ordered ferroelectric state ($\boldsymbol{\delta}\neq0$) and the dipole-isotropic paraelectric state ($\boldsymbol{\delta}=0$) is, as shown in Appendix \ref{appendixI}-(ii),
\begin{eqnarray}
	\Delta S(\boldsymbol{\delta})=-Nd\left[\frac12|\log(1-\bar{u}^2(\delta))|\right]+o(d).	\label{ent_delta}
\end{eqnarray}
It follows
\begin{equation}
	\Delta F_{\mathrm{ent}}(\boldsymbol{\delta})= - \frac{1}{\beta}\ \Delta S(\boldsymbol{\delta})>0. \label{DeltaFent}
\end{equation}
As anticipated in Sec. \ref{par3_1}, the entropy loss associated with dipolar ordering scales as $d$.
The excess free-energy difference between ferroelectric and  paraelectric state in the large-$d$ limit is, as derived in Appendix \ref{appendixI}-(iv),
\begin{align}
	\Delta\mathcal{F}(\rho,\boldsymbol{\delta})=\frac{N\rho}{\beta}B_2^{HS}d\int_{-\infty}^{\infty} dh e^h e^{-\beta {v}_0(h)} f_{\delta}(h); \nonumber \\
	f_{\delta}(h)=-\bar{f}_{w_p}\big(\bar{u}^2(\delta),h\big) \label{deltaFex}
\end{align}
with $\bar u(\delta)$ given in Eq. \ref{pdelta}. Because $\bar f_{w_p}(\hat d_i \cdot \hat d_j,h) \geq 0$, $f_\delta(\rho,\delta)<0$, $\forall t$, see also Fig. \ref{fig3}. As a consequence, a finite value of $\delta$ lowers $\mathcal F$. 
Observe that, according to Eqs. \ref{ent_delta}-\ref{deltaFex}, both $\Delta F_{\mathrm{ent}}(\boldsymbol{\delta})$ and $\Delta \mathcal{F}(\rho,\boldsymbol{\delta})$ scale as $O(Nd)$. While the former is strictly positive and therefore penalizes orientational ordering, the latter is negative and favors the onset of ferroelectric order. The dependence of $\Delta\mathcal{F}$ on the density has been made explicit. Moreover, as follows from Eq. \ref{deltaFex}, $\Delta\mathcal{F}$ also depends on temperature.
The thermodynamic parameters can therefore drive the quantity $\Delta F_{\mathrm{ent}}(\boldsymbol{\delta})+\Delta\mathcal{F}(\rho, \boldsymbol{\delta})$ from positive to negative values, inducing the ferroelectric phase transition.  
A Taylor expansion of the two contributions for small $\delta$ yields a Landau-like free-energy expression in terms of the ferroelectric order parameter. Similar developments are presented in Ref. \cite{Izzo}, where the interplay between ferroelectric and liquid-liquid phase transitions is also discussed in detail, and are beyond the scope of the present study. As  emphasized in Ref. \cite{Izzo}, the emergence of a ferroelectric phase transition relies on the negative sign of $\Delta \mathcal{F}(\boldsymbol{\delta})$. The present analysis identifies the physical origin of this negative contribution, showing it to be a direct consequence of the presence of annealed positional disorder.

\subsection{Short-Range Annealed-Averaged Dipolar Interaction: Hindered Dipolar Rotation and Pair Correlations} \label{par3_4}
Once the explicit expression of $ \bar{f}_{w_p}(r, \hat{d}_i \cdot \hat{d}_j)$ is known, the effective potential $\bar{W}_{d \rightarrow \infty}$ in Eq. \ref{Wbar_dinf} can be obtained in closed form,
\begin{widetext}
\begin{eqnarray}
	\bar{W}_{d \rightarrow \infty}=\frac{1}{\beta} \big[t\hat{d}_i \cdot \hat{d}_j+\frac{1}{2}\log[1-2 t\hat{d}_i \cdot \hat{d}_j -t^2(1-{\hat{d}_i \cdot \hat{d}_j}^2)] \big], \ \ \ \ t=\beta p^2 \theta(h)e^{- h}. \label{wtildep}
\end{eqnarray} 
\end{widetext}
In the following, the effective interaction $\bar W$ is shown to be shorter-ranged than the bare dipolar potential $w_p$, as a consequence of the screening induced by annealed positional disorder. This mechanism mirrors the screening arising from orientational averaging of freely rotating dipoles, which leads to the Keesom interaction.
From Eqs. \ref{Wbar_dinf} and \ref{fbardinf_1}, Eq. \ref{wtildep} can be rewritten as
\begin{equation}
	\bar{W}_{d \rightarrow \infty} = -\frac{1}{\beta}K_{\tilde{Z}}(t), \label{WbarKz}
\end{equation}
where $K_{\tilde{Z}}(t)=\log M_{\tilde{Z}}(t)$ is the cumulant generating function associated with the random variable $\tilde Z$.
Using the cumulant expansion
\begin{equation}
	K_{\tilde Z}(t)
	=
	\sum_{n=1}^{\infty}\frac{1}{n!}\,
	k_{\tilde Z}^{(n)}\,t^n,
	\qquad
	k_{\tilde Z}^{(n)}
	=
	\left.
	\frac{d^n}{dt^n}K_{\tilde Z}(t)
	\right|_{t=0},
\end{equation}
Eq. \ref{WbarKz} becomes
\begin{eqnarray}
	\bar{W}_{d \rightarrow \infty}=-\frac{1}{\beta} \sum_{n=2}^{\infty}\frac{1}{n!}(\beta \bar{p}^2)^n \theta(h)e^{-nh} k_{\tilde{Z}}^{(n)}  . \label{K}
\end{eqnarray}
The leading nonvanishing contribution in Eq. \ref{K} arises at $n=2$, since $k_{\tilde Z}^{(1)}=\left.\frac{d}{dt}\log \langle e^{t\tilde Z}\rangle_{\hat r_{ij}}\right|_{t=0}=\left.\langle \tilde Z e^{t\tilde Z}\rangle_{\hat r_{ij}}/\langle e^{t\tilde Z}\rangle_{\hat r_{ij}}\right|_{t=0}=\langle \tilde Z\rangle_{\hat r_{ij}}=0$. $\bar W$ thus scales as $e^{-2h}+o(e^{-2h})$, which, upon reverting from $h$ to the radial coordinate $r$, yields 
\begin{equation}
	\bar W_{d \rightarrow \infty}(r)\propto r^{-2d}+o(r^{-2d}).
\end{equation}
Therefore, $\bar W_{d \rightarrow \infty}$ is shorter-ranged than the bare dipolar interaction.

\begin{figure*}[t]
	\centering
	\includegraphics[width=0.85\linewidth]{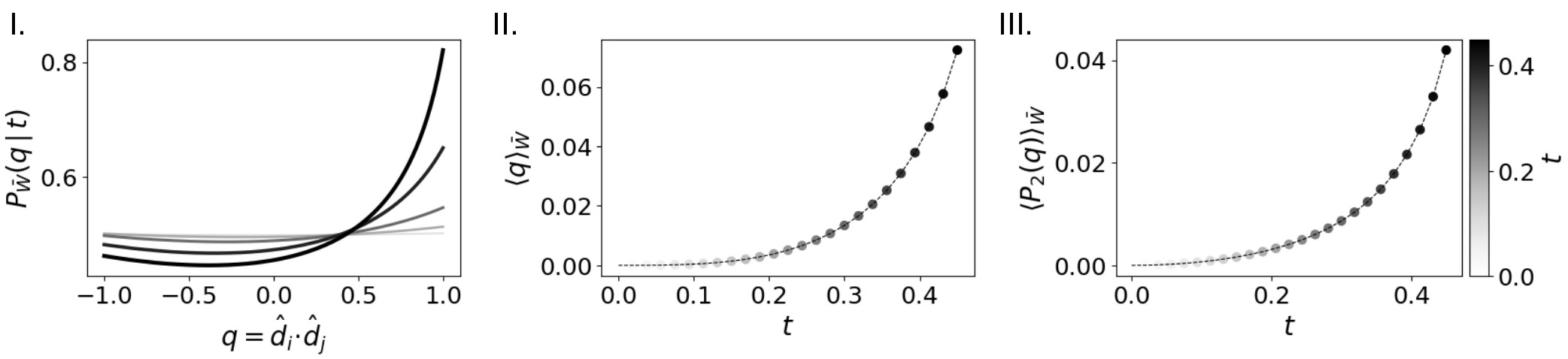}
	\caption{Boltzmann probability distribution associated to the effective potential $\bar W$, $P_{\bar W}(q|t)$ (\textit{Panel I}) together with the corresponding first spherical-harmonic moment, $\langle q \rangle_{\bar W}$ (\textit{Panel II}), and second spherical-harmonic moment $\langle P_2(q)\rangle_{\bar W}$. All quantities are shown for different values of the parameter $t$. The second-order Legendre polynomial is defined as $P_2(q)=dq^2-1$, evaluated here for $d=3$.}
	\label{fig4}
\end{figure*}

$\bar{W}_{d \rightarrow \infty}$ has a ferroelectric-like character, as shown in Sec. \ref{par3_3}. Following the discussion in Sec. \ref{intro} this is expected to result in a positive value of $\langle \hat{d}_i \cdot \hat{d}_j \rangle$ even in the paraelectric phase. Eq. \ref{W} can be recast as
\begin{align}
	\langle \hat d_{i}\cdot\hat d_{j}\rangle_{\boldsymbol{\delta}=0}
	&= \frac{1}{Z_0} \int_{-\infty}^{\infty} d h \ e^h e^{-\beta  v_0(h)}
	\nonumber\\
	&\quad\times
	\int_{-1}^{1} dq\ q\, p_d(q)\ e^{-\beta \bar W_{d \rightarrow \infty}(h,q)}, 
	\label{Wu}
\end{align}
where 
\begin{equation}
	q=\hat d_i\cdot\hat d_j,
\end{equation} 
and $Z_0=\int_{-\infty}^{\infty} d h \ e^h e^{-\beta  v_0(h)}\int_{-1}^{1} dq\ p_d(q)\
e^{-\beta \bar W(\mathbf h,q)}$. The probability density of $q$ for two independent unit vectors $\hat{d}_{i(j)}$ uniformly distributed in $\mathbb R^d$ is
\begin{equation}
	p_d(q)=\frac{\Omega_{d-1}}{\Omega_{d}}(1-q^2)^{\frac{d-3}{2}}, \label{pdq}
\end{equation}
as obtained by the same argument used in Appendix \ref{appendixI}-(i).
Since $\bar W_{d \rightarrow \infty}$  remains finite, while $p_d(q)$ becomes increasingly peaked around $q=0$ for large $d$, Eq. \ref{Wu} implies $\lim_{d\to\infty}\langle \hat d_i\cdot\hat d_j\rangle=0$. 
Although exact in the limit $d \rightarrow \infty$, this result does not necessarily capture the behavior at large but finite dimensionality when approaching the $d\to\infty$ limit, where the functional form of $\bar W(r,q)$ may still significantly affect dipolar correlations. Increasing $d$ primarily suppresses the magnitude of $\langle \hat d_i\cdot\hat d_j\rangle$ through $p_d(q)$, without altering the ferroelectric character of $\bar W_d$. To isolate the effect of $\bar W_{d \rightarrow \infty}$ on hindered dipolar rotations, the case of a uniform angular measure $p_d(q)$ in Eq.~\ref{Wu} is analyzed below. 
Fig. \ref{fig4} reports: (i) the canonical probability distribution of $q=\hat d_i\cdot\hat d_j$ associated with $\bar W_{d \rightarrow \infty}$, $P_{\bar W}(q\,|\,t)=e^{-\beta \bar W_{d \rightarrow \infty}}/Z_{\bar W}$ with $Z_{\bar W}=\int_{-1}^1dq \ e^{-\beta \bar W_{d \rightarrow \infty}(h,q)}$ ; (ii) the corresponding first spherical-harmonic moment ($\ell=1$), $\langle \hat q \rangle_{\bar W}$, measuring local dipolar alignment; and (iii) the second spherical-harmonic moment ($\ell=2$), $\langle P_2(q)\rangle_{\bar W}$,  quantifying local quadrupolar (nematic-like) ordering. All quantities refer to the paraelectric phase, since Eq. \ref{W} implicitly assumes $\zeta(\hat{d})$ to be uniform over the solid angle.
Fig. \ref{fig4} confirms that $\langle q\rangle_{\bar W}>0$, $\forall t$. This follows from the ferroelectric character of $\bar W_{d \rightarrow \infty}$, which favors parallel dipolar alignment, as directly illustrated by Panel I of Fig. \ref{fig4}, where $P_{\bar W}(q\,|\,t)$ is biased toward positive values of $q$.
These results show that a positive value of the dipolar correlation $\langle \hat d_i\cdot\hat d_j\rangle$ in the paraelectric phase, readily accessible in numerical simulations, already signal the tendency of the system to develop a ferroelectric phase transition.
If the first moment of $P_{\bar W}(q|t)$ already captures the qualitative ferroelectric-like character of $\bar W$, higher-order moments are expected to provide a more quantitative characterization. They could in principle be exploited to reconstruct $P_{\bar W}(q|t)$. In numerical simulations, this can be achieved by computing a finite set of moments and solving the associated inverse problem, for instance via maximum-entropy methods \cite{Mead}. 
By exploiting the equivalence, pointed out in Sec. \ref{intro}, between $P_{\bar W}(q|t)$ and the pair correlation function $g^{(2)}$ entering the DFT expression of $\mathcal{F}$, see e.g. Eq. \ref{F_DFT}, this approach can thus provide direct numerical access to $g^{(2)}$. 
This, in turn, would enable identification of the onset of the ferroelectric phase transition and of the critical temperature from numerical simulations performed entirely within the paraelectric phase, providing an alternative to numerical simulation studies across the putative critical temperature, which rely on demanding finite-size scaling analyses. 
A different strategy is presented in Ref. \cite{Sammuller}, where the DFT free-energy functional is learned directly from numerical simulation data using a neural-network approach.

Following Eq. \ref{W1}, in the ferroelectric phase and in the limit $d \rightarrow \infty$, it is
\begin{align}
	\langle \hat d_{i}\cdot\hat d_{j}\rangle_{\boldsymbol{\delta}}
	&=
	\frac{1}{Z_{\boldsymbol{\delta}}}
	\int_{-\infty}^{\infty}d h \ e^he^{-\beta v_0(h)}
	\int d\hat d_i d \hat d_j \,
	\zeta(\hat d_i) \zeta(\hat d_j)
	\nonumber\\
	&\quad\times
	\hat d_i \cdot \hat d_j \ 
	e^{-\beta \bar W_{d \rightarrow \infty}(h,\hat d_i \cdot \hat d_j)},
	\label{Wu_ferro}	
\end{align}
where $Z_{\boldsymbol{\delta}}=\int_{-\infty}^{\infty} d h \ e^he^{-\beta v_0(h)} \int d\hat d_i d \hat d_j \  \zeta(\hat d_i) \zeta(\hat d_j)  e^{-\beta \bar W_{d \rightarrow \infty}}$ and $\zeta(\hat d)$ is given in Eq. \ref{zeta} with $\boldsymbol{\delta} \neq 0$.
The same steps used in Appendix \ref{appendixI}-(iv), based on the Laplace saddle-point method, readily show that
\begin{eqnarray}
	\langle \hat d_i \cdot \hat d_j\rangle_{\boldsymbol \delta}=\bar u^2(\delta)=\boldsymbol{\bar{p}}^2, \label{corr_ferro}
\end{eqnarray}
where $\bar u(\delta)$ is given in Eq. \ref{pdelta}.
Since, as shown above, in the limit $d\to\infty$, $\langle \hat d_i \cdot \hat d_j\rangle_{\boldsymbol{\delta}=0}=0$
in the paraelectric phase, Eq. \ref{corr_ferro} suggests that the dipolar correlation $\langle \hat d_i \cdot \hat d_j \rangle$ measured in numerical simulations may provide an indicator of the onset of the ferroelectric phase transition. 
In finite $d$, however, $\langle \hat d_i \cdot \hat d_j\rangle$ retains a local contribution even within the paraelectric phase, which becomes progressively suppressed as $d\to\infty$, as emphasized above.
If using $\langle \hat d_i \cdot \hat d_j\rangle$ as a diagnostic of
ferroelectric ordering, it is important to recall that the hallmark of the ferroelectric phase transition is the emergence of a macroscopic polarization in the ordered phase. Eq. \ref{P} indeed implies that $\lim_{N \rightarrow \infty}\frac{1}{N^2} \sum_{i,j=1}^N \langle \hat{d}_i \cdot \hat{d}_j \rangle$ is $O(1)$ in the ferroelectric phase, while it vanishes in the paraelectric phase. Distinguishing between the ferroelectric and paraelectric phases therefore requires finite-size scaling analyses in numerical simulations to determine whether $\langle \hat d_i\cdot\hat d_j\rangle$ exhibits different scaling behavior in the two phases.

\begin{figure}[t]
	\centering
	\includegraphics[width=0.65\linewidth]{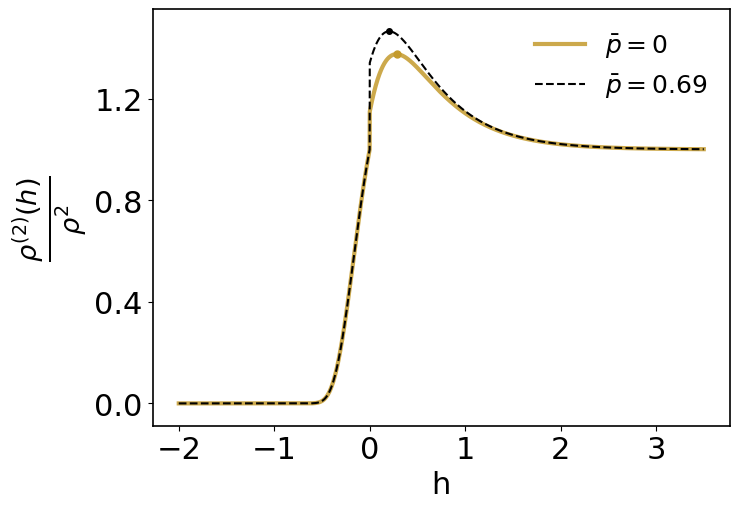}
	\caption{Radial pair correlation marginalized over dipolar degrees of freedom, $\frac{\rho^{(2)}}{\rho^2}$, in the paraelectric ($\delta=0$, solid line) and ferroelectric ($\delta \neq 0$, dashed line) phases.}
	\label{fig5}
\end{figure}

The emergence of ferroelectric order also leaves a clear signature on the radial pair correlation function marginalized over the dipolar degrees of freedom, $\rho^{(2)}(r)$, which can therefore serve as an indicator of the onset of the ferroelectric phase transition. Starting from Eq. \ref{pair_lowrho}, using the definition of $\bar W_{d\to\infty}$ in Eq.~\ref{Wbar_dinf}, the one-particle density in Eq. \ref{rho_om} and the scaling variable $h$, one obtains
\begin{align}
	\frac{\rho^{(2)}(h)}{\rho^2}
	&=
	e^{-\beta v_0(h)}
	\int d\hat d_i\,d\hat d_j\,
	\zeta(\hat d_i)\,\zeta(\hat d_j)
	\nonumber\\
	&\quad\times
	e^{-\beta \bar W_{d \rightarrow \infty}(h,\hat d_i \cdot \hat d_j)}.
	\label{rho2ferro}
\end{align}
For the ansatz $\zeta(\hat d)$ in Eq. \ref{zeta}, Eq. \ref{rho2ferro} reduces to
\begin{eqnarray}
	\frac{\rho^{(2)}_{\boldsymbol{\delta}}(h)}{\rho^2}=e^{-\beta v_0(h)} \ e^{-\beta \bar W_{d \rightarrow \infty}(h,\bar u^2(\delta))},  \label{rho2delta}
\end{eqnarray} 
where $\bar u(\delta)$ is introduced in Eq. \ref{pdelta} and the Laplace method has been exploited. The derivation follows the same steps used in Appendix \ref{appendixI}-(iv), in particular Eqs. \ref{uijdef}-\ref{ubar2}. Eq. \ref{rho2delta} applies to both the ferroelectric and paraelectric phases. In the latter case, $\delta=0$, which implies $\bar p=\bar u(\delta)=0$.
Fig. \ref{fig5} shows the radial pair correlation $\rho^{(2)}_{\boldsymbol{\delta}}(h)/\rho^2$ in the paraelectric ($\bar{p}=0$, full line) and ferroelectric phase ($\bar{p}=0.69$, dashed line).  Lennard-Jones potential is chosen for the reference potential, so that $ v_0(h)=E_0\big(e^{-\nu h}-e^{-\nu h/2}\big)$ \cite{Parisi}, with $\nu=4$. 
Compared to the paraelectric phase, the characteristic peak of the pair correlation function in the ferroelectric phase is enhanced and slightly shifted toward smaller values of $h$, corresponding to reduced characteristic interparticle distances. 
This indicates that ferroelectric ordering promotes not only orientational correlations, but also local positional ordering, increasing the probability of finding particles at the same characteristic distance associated with the peak. The shift towards shorter distances can be understood intuitively by noting that, within a mean-field picture in which $p\hat d_i=p\hat d_j=\boldsymbol{\bar p}$, the annealed averaged dipolar potential $\bar W_{d \rightarrow \infty}$ is attractive as a function of the interparticle distance $r$.
This result is noteworthy because $\rho^{(2)}(r)/\rho^2$ coincides with the radial pair correlation function $g^{(2)}(r)$ measured in numerical simulations when dipolar degrees of freedom are not explicitly resolved.
In real fluids, $g^{(2)}(r)$ typically exhibits multiple coordination shells. Nevertheless, this simplified scenario highlights an important mechanism: dipolar ordering alone can promote local spatial ordering, leaving clear signatures in the radial correlation function $g^{(2)}(r)$. This observation may provide a useful perspective for interpreting changes observed in $g^{(2)}(r)$ between the high-density and low-density phases of supercooled water \cite{wikfeldt}. Moreover, it offers further consistency with associating a paraelectric character to the high-density and a ferroelectric character to the low-density phase.

\section{From infinite to finite dimensions} \label{cap4}
To assess the relevance of the mechanism identified in Secs. \ref{par3_1}-\ref{par3_3} for real systems, it is important to clarify whether it persists at finite $d$. The key issue is whether an annealed average over $\hat r_{ij}$ of a suitably renormalized dipolar two-body potential, $W_d$, incorporating many-body contributions that are no longer negligible at finite $d$, generates an effective ferroelectric-like interaction. The many-body effects at finite-$d$ can be accounted for through the renormalized dipolar potential defined within the optimized cluster expansion, introduced in Sec. \ref{par4_2}. Sec. \ref{par4_3} then shows that its annealed average over $\hat r_{ij}$ indeed generates a ferroelectric-like dipolar interaction. 
In the present framework, the pair correlation function entering the DFT functional is related to the effective interaction through
\begin{equation}
	V(r, \hat r_{ij},\hat d_i,\hat d_j)=-\frac{1}{\beta}\log g^{(2)}(r,\hat r_{ij},\hat d_i,\hat d_j), \label{V_g}
\end{equation}
up to a physically irrelevant additive constant. Consistently with Sec. \ref{intro}, decompose $V=V_0+W_d$, where $V_0$ is the isotropic contribution of the reference system to $V$ and $W_d$ the effective two-body dipolar interaction, which is a functional of $w_p$. This decomposition holds within the optimized cluster expansion scheme adopted here to derive $V$.
Introducing the perturbation parameter $\lambda$ as in Eq. \ref{F_DFT},
\begin{equation}
	V=V_0+\lambda W_d.
\end{equation}
Assuming that $W_d=W_d[w_p]$ is analytic around $w_p=0$, so that
Eq. \ref{averagerij} applies, one obtains
\begin{widetext}
	\begin{align}
		\lim_{R_c \rightarrow \infty} \mathcal{F}[\zeta]
		&=
		-C^d_{W}p^2f_d(\epsilon)\frac{\Omega_d}{d}
		\int  dr \ r^{d-1}e^{-\beta V_0(r)}
		d\hat d_i d \hat d_j \ 
		\hat d_i \cdot \hat{d}_j
		\nonumber\\
		&\quad+
		\frac{1}{2}N\rho \Omega_d
		\int  dr \ r^{d-1}e^{-\beta V_0(r)}
		d \hat d_i d \hat{d}_j
		\zeta(\hat d_i)
		\langle e^{-\beta W_d(r, \hat r_{ij},\hat d_i,\hat d_j)}-1\rangle_{\hat{r}_{ij}}
		\zeta(\hat d_j).
		\label{FdRd}
	\end{align} 
\end{widetext}
The constant $C_W^d$ originates from the component of $W_d[w_p]$ linear in $w_p$. For $C_W^d > 0$, the first term on the right-hand side of Eq. \ref{FdRd}, favors ferroelectric ordering. If negative, it suppresses it.

In the Supplementary Material, an analytical expression for the excess free energy at finite $d$, resulting from the second-order truncation of the virial expansion, is derived, and shown to be minimized by ferroelectric order for all finite $d \geq 3$.
However, truncating the virial expansion at second order is generally insufficient to accurately describe real systems. The renormalized potential within the optimized cluster expansion is then introduced below. While effectively incorporating many-body contributions, this construction still allows the free-energy functional $\mathcal F$ to retain the same formal structure as in the second-order virial truncation.

\subsection{Effective Two-Body Dipolar Interaction as Perturbative Potential in the Optimized Cluster Expansion} \label{par4_2}

By decomposing the interaction potential into a hard-core reference contribution and a perturbative term, the optimized cluster expansion for classical fluids \cite{Hansen, Chandler0} provides an approximate expression for $W_d$. Within this framework, $W_d$ coincides, up to a sign and a factor $\beta$, with the so-called renormalized potential $\mathcal C_p$ obtained from a resummation of the virial series. Its diagrammatic representation involves generalized chains in which density vertices are replaced by hypervertices, functionals of the pair correlation function of the reference system, $g_0(\mathbf r_i,\mathbf r_j)$. Physically, the renormalized potential describes an effective interaction in which the perturbation potential is screened by the local structure imposed on the fluid by the reference potential \cite{Hansen}. Only the aspects relevant to the present treatment are summarized below.  Further details can be found in Refs. \cite{Hansen,Chandler0}. For simplicity, when referring to a generic case, the dipolar degrees of freedom are neglected and a potential depending only on the position vector is considered.

In the optimized cluster expansion the pair density entering the DFT expression of $\mathcal{F}$, is given by the following exponential approximation \cite{Chandler0, Hansen} 
\begin{eqnarray}
	\rho^{(2)}(\mathbf{r}_i,\mathbf{r}_j)=\tilde{\rho}(\mathbf{r}_i)g_0(\mathbf{r}_i,\mathbf{r}_j)e^{\mathcal{C}_p(\mathbf{r}_i,\mathbf{r}_j,\beta)}\tilde{\rho}(\mathbf{r}_j). \label{expapprox}
\end{eqnarray}
Eq. \ref{expapprox} is asymptotically correct \cite{Chandler0} in each of the four limits: (i) the low-density limit, where it recovers the virial expansion with truncation at second order \cite{Hansen}; (ii) the high-density limit; (iii) the high-temperature limit or weak coupling; and (iv) the $\gamma \rightarrow 0$ limit, where $\gamma^{-1}$ is the range of the perturbation potential in the $\gamma$ expansion, i.e., the long-range limit where the mean field becomes exact.
In classical DFT, treating $\mathcal{C}_p$ as a perturbative interaction, and defining $\rho^{(2)}(\mathbf{r}_i,\mathbf{r}_j,\lambda)=\tilde{\rho}(\mathbf{r}_i)g_0(\mathbf{r}_i,\mathbf{r}_j)e^{\lambda\mathcal{C}_p(\mathbf{r}_i,\mathbf{r}_j,\beta)}\tilde{\rho}(\mathbf{r}_j)$, following Eq. \ref{expapprox}, substitution into Eq. \ref{F_DFT} and integration over $\lambda$ yield  
\begin{eqnarray}
	\mathcal{F}[\tilde{\rho}]=-\frac{1}{2} \int d\textbf{r}_i d\textbf{r}_j \tilde{\rho}(\mathbf{r}_i)g_0(\mathbf{r}_i,\mathbf{r}_j) [e^{\mathcal{C}_p(\mathbf{r}_i,\mathbf{r}_j,\beta)}-1 ] \tilde{\rho}(\mathbf{r}_j).	\ \label{F_DFTC}
\end{eqnarray}   
Defining $V_0(\textbf{r}_i,\textbf{r}_j)=-\frac{1}{\beta} \log g_0(\mathbf{r}_i,\mathbf{r}_j)$ Eq. \ref{F_DFTC} can be recognized as the finite-$d$ counterpart of the virial expansion truncated at second order, exact in the limit $d\to\infty$. In this representation, the bare perturbative interaction $w_p$ is replaced by the renormalized potential $-\frac{1}{\beta}\mathcal C_p$, while the factor $e^{-\beta v_0(r)}$ is replaced, through topological reduction \cite{Chandler0,Hansen,Morita}, by $g_0(\mathbf r_i,\mathbf r_j)$.

The explicit expression for the potential $\mathcal{C}_p$  follows from a diagrammatic resummation of the virial series. The total Mayer function in Eq. \ref{Mayer_vowp}, representing the elementary bond of the virial expansion, can be rephrased by replacing the factor $e^{-\beta v_0(r)}$ with $g_0(\mathbf r_i,\mathbf r_j)$ and expanding $f_{w_p}$ in powers of $w_p$, yielding
\begin{align}
	\tilde{f}(\mathbf{r}_i,\mathbf{r}_j)
	&=
	h_0(\mathbf{r}_i,\mathbf{r}_j)
	+
	(1+h_0(\mathbf{r}_i,\mathbf{r}_j))
	\nonumber\\
	&\quad\times
	\sum_{n=1}^{\infty}
	(-1)^n \beta^n
	[w_p(\mathbf{r}_i,\mathbf{r}_j)]^n,
	\label{Mayertop}
\end{align}
The tilde indicates that the Mayer function has been modified via topological reduction and $h_0(\mathbf{r}_i,\mathbf{r}_j)=g_0(\mathbf{r}_i,\mathbf{r}_j)-1$.
The bond $\tilde f$ appearing in the virial diagrams can be decomposed into three contributions: a single $h_0$ bond, any number of powers of $w_p$ bonds, any number of powers of $w_p$ bonds times $h_0$ bond. The renormalized potential is obtained by resumming chains of $w_p$ bonds in the virial series \cite{Chandler0, Hansen}. In summary, one obtains
\begin{eqnarray}
	\rho^2\mathcal{C}_p(\mathbf{r}_i,\mathbf{r}_j,\beta)=\rho^2\sum_{n=1}^{\infty}\mathcal{C}_p^{(n)}(\mathbf{r}_i,\mathbf{r}_j,\beta). \label{C}
\end{eqnarray} 
Each term $\mathcal{C}_p^{(n)}(\mathbf{r}_i,\mathbf{r}_j, \beta)$ corresponds to a convolution integral which, in Fourier space with conjugate variable $\mathbf k$, reads
\begin{eqnarray}
	\rho^2\tilde{\mathcal{C}}_p^{(n)}(\mathbf{k},\beta)=(-1)^n(\beta)^n[ \tilde{w}_p(\mathbf{k})\tilde{\Sigma}_0(\mathbf{k})]^n\tilde{\Sigma}_0(\mathbf{k}), \label{Cn}
\end{eqnarray}
where $\tilde{\mathcal{C}}^{(n)}(\mathbf{k},\beta)$, $\tilde{w}_p(\mathbf{k})$, and $\tilde{\Sigma}_0(\mathbf{k})$ denote the Fourier transforms respectively of $\mathcal{C}_p^{(n)}(\mathbf{r}_i,\mathbf{r}_j)$, $w_p(\mathbf{r}_i,\mathbf{r}_j)$, and of the so-called hypervertex function. The latter is defined as
\begin{eqnarray}
	\Sigma_0(\mathbf{r}_i,\mathbf{r}_j)=\rho \delta(\mathbf{r}_i,\mathbf{r}_j)+\rho^2 h_0(\mathbf{r}_i,\mathbf{r}_j), \label{sigma}
\end{eqnarray}
and generalizes the elementary vertex $\rho \delta(\mathbf r_i,\mathbf r_j)$ appearing in the standard diagrammatic virial expansion by incorporating the local structure of the reference fluid through $h_0$.
For example, for $n=1$, the corresponding expression in real space reads 
\begin{align}
	\rho^2 \mathcal{C}_p^{(1)}(\mathbf{r}_i,\mathbf{r}_j,\beta)
	&=
	-\beta \int \int
	\Sigma_0(\mathbf{r}_i,\mathbf{r}_k)
	w_p(\mathbf{r}_k,\mathbf{r}_l)
	\nonumber\\
	&\quad\times
	\Sigma_0(\mathbf{r}_l,\mathbf{r}_j)
	d \mathbf{r}_k d\mathbf{r}_l.
	\label{C1}
\end{align}
As follows from Eqs. \ref{C}-\ref{C1}, the renormalized potential can be read as an effective interaction in which the bare perturbation is screened by the local structure generated in the fluid by the reference potential \cite{Hansen}.

\subsection{Annealed Averaging of the Renormalized Potential\\
	over Positional Disorder and Ferroelectricity in Finite Dimensions} \label{par4_3}
Eq. \ref{F_DFTC}, which applies at finite $d$, can be recast in a form analogous to Eq. \ref{Fexdinf}, valid in the limit $d\to\infty$, as
\begin{align}
	\mathcal{F}[\tilde{\rho}]
	&=
	-\frac{N\rho}{2}\Omega_d
	\int r^{d-1} d r \ g_0(r)
	d\hat{d}_i d\hat{d}_j\,
	\zeta (\hat{d}_i)
	\nonumber\\
	&\quad\times
	\langle
	f_{\mathcal{C}_p}(r,\hat{r}_{ij},\hat{d}_i,\hat{d}_j)
	\rangle_{\hat{r}_{ij}}
	\zeta (\hat{d}_j).
	\label{Fexd}
\end{align}
The modified Mayer function associated with the renormalized dipolar potential is
\begin{equation}
	f_{\mathcal{C}_p}(r, \hat{r}_{ij},\hat{d}_i, \hat{d}_j)=e^{\mathcal{C}_p}-1. \label{fcp}
\end{equation}
Since $
\langle
\mathcal C_p^{(1)}
\rangle_{\hat r_{ij}}
=0
$, as follows from Eq. \ref{C1} and the isotropy of $\Sigma_0$, which preserves under convolution the angular dependence inherited from $v_p(\hat r_{ij})$, the leading non-vanishing isotropic contribution arise at order $w_p^2$. Under the approximation that the dominant isotropic sector is given by the $\ell=0$ projection of $w_p^2$, weighted by $\Sigma_0$, whose positivity is proven in Appendix \ref{appendixII}-(i), this contribution is non-negative. Therefore, by Jensen's inequality, one can safely assume $
\langle
e^{\mathcal C_p}-1
\rangle_{\hat r_{ij}}
\ge0
$. Consequently, minimizing $\mathcal F$ in Eq.~\ref{Fexd} is equivalent to maximizing $
\langle
f_{\mathcal C_p}
\rangle_{\hat r_{ij}}
$.

To evaluate $\langle f_{\mathcal{C}_p} \rangle_{\hat{r}_{ij}}$ in Eq. \ref{Fexd}, $e^{\mathcal{C}_p}-1$  is expanded in powers of $\mathcal C_p$ in real space. Using Eq. \ref{C}, one obtains
\begin{eqnarray}
	\langle e^{\mathcal{C}_p}-1 \rangle_{\hat r_{ij}}
	=
	\sum_{n=1}^{\infty}\sum_{l=1}^n
	\frac{1}{l!}
	\sum_{\substack{n_1+\cdots+n_l=n\\ n_i\ge1}} \langle
	\mathcal{C}_p^{(n_1)}\cdots\mathcal{C}_p^{(n_l)} \rangle_{\hat r_{ij}}. \ \
	\label{expCseries}
\end{eqnarray}
It is convenient to decompose $w_p$ as in the following. Let
\begin{equation}
	w_p(\mathbf r_{ij},\hat d_i,\hat d_j)
	=
	v_p(\hat r_{ij},\hat d_i,\hat d_j)\,q_p(r), \label{wp_fat}
\end{equation}
with
\begin{align}
	v_p(\hat r_{ij},\hat d_i,\hat d_j)
	&=
	d(\hat d_i\!\cdot\!\hat r_{ij})
	(\hat d_j\!\cdot\!\hat r_{ij})
	-
	\hat d_i\!\cdot\!\hat d_j,
	\nonumber\\
	q_p(r)
	&=
	-p^2\Big(\frac{l}{r}\Big)^{ d}\theta(r-l),
\end{align}
where $v_p$ and $q_p$ denote respectively the angular and isotropic radial contributions.
Eq. \ref{Cn} then becomes
\begin{align}
	\rho^2\tilde{\mathcal{C}}_p^{(n)}
	(\textbf{k}_{ij},\hat d_i,\hat d_j,\beta)
	&=
	(\beta)^n
	[v_p(\hat k_{ij},\hat d_i,\hat d_j)]^n
	\nonumber\\
	&\quad\times
	[-\tilde q_p(k)\tilde\Sigma_0(k)]^n
	\tilde\Sigma_0(k),
	\label{Cn_iso}
\end{align}
where $\mathbf k_{ij}=k\hat k_{ij}$ and $\tilde q_p(k)$ denotes the Fourier transform of $q_p(r)$.
Note that, as emphasized in Eq. \ref{Cn_iso}, the Fourier transform
\begin{equation}
	\tilde w_p(\mathbf k_{ij},\hat d_i,\hat d_j)
	=
	\int d\mathbf r_{ij}\,
	e^{-i\mathbf k_{ij} \cdot\mathbf r_{ij}}
	w_p(\mathbf r_{ij},\hat d_i,\hat d_j)
\end{equation}
preserves the angular structure of the dipolar interaction: the dependence on $\hat r_{ij}$ is mapped onto the corresponding dependence on $\hat k_{ij}$. This follows by expanding both $v_p(\hat r_{ij},\hat d_i,\hat d_j)$ and $e^{-i\mathbf k_{ij} \cdot\mathbf r_{ij}}$ in spherical harmonics. The integration over $\hat r_{ij}$ selects the same $\ell=2$ harmonic component appearing in the expansion of $v_p(\hat r_{ij},\hat d_i,\hat d_j)$, so that only the radial part is effectively transformed trough a spherical-Bessel transform with $\ell=2$.

Define
\begin{equation}
	\mathcal C_p^{(n_1,\ldots,n_l)}
	(\mathbf r_{ij},\hat d_i,\hat d_j)
	\equiv
	\prod_{a=1}^{l}
	\mathcal C_p^{(n_a)}
	(\mathbf r_{ij},\hat d_i,\hat d_j).
\end{equation}
Expanding in spherical harmonics,
\begin{align}
	\mathcal C_p^{(n_1,\ldots,n_l)}
	(\mathbf r_{ij},\hat d_i,\hat d_j)
	&=
	\sum_{\ell,\mu}
	\mathcal Q_{p,\ell}^{(n_1,\ldots,n_l)}(r)\,
	\nonumber\\
	&\quad\times
	a_{\ell\mu}^{(n_1,\ldots,n_l)}
	(\hat d_i,\hat d_j)
	Y_{\ell\mu}(\hat r_{ij}),
\end{align}
the Fourier transform acts diagonally on each angular sector:
\begin{align}
	\widetilde{\mathcal C}_p^{(n_1,\ldots,n_l)}
	(\mathbf k_{ij},\hat d_i,\hat d_j)
	&=
	\sum_{\ell,\mu}
	\widetilde{\mathcal Q}_{p,\ell}^{(n_1,\ldots,n_l)}(k)\,
	\nonumber\\
	&\quad\times
	a_{\ell\mu}^{(n_1,\ldots,n_l)}
	(\hat d_i,\hat d_j)
	Y_{\ell\mu}(\hat k_{ij}),
\end{align}
where
\begin{align}
	\widetilde{\mathcal Q}_{p,\ell}^{(n_1,\ldots,n_l)}(k)
	&=
	(2\pi)^{d/2}
	i^{-\ell}
	k^{1-\frac d2}
	\int_0^\infty dr\,
	r^{\frac d2}
	\nonumber\\
	&\quad\times
	\mathcal Q_{p,\ell}^{(n_1,\ldots,n_l)}(r)
	J_{\ell+\frac d2-1}(kr).
\end{align}
$J_{\nu}(x)$ is the Bessel function of the first kind of order $\nu$.
A completely analogous inverse relation holds for $\mathcal Q_{p,\ell}^{(n_1,\ldots,n_l)}(r)$.
Taking the angular averages
$
\langle\ \rangle_{\hat r_{ij}}
$
and
$
\langle\ \rangle_{\hat k_{ij}}
$
selects only the isotropic sector $\ell=0$, yielding
\begin{widetext}
\begin{equation}
	\langle
	\widetilde{\mathcal C}_p^{(n_1,\ldots,n_l)}
	\rangle_{\hat k_{ij}}
	=
	\widetilde{\mathcal Q}_{p,0}^{(n_1,\ldots,n_l)}(k)\,
	a_{00}^{(n_1,\ldots,n_l)},
	\qquad
	\langle
	\mathcal C_p^{(n_1,\ldots,n_l)}
	\rangle_{\hat r_{ij}}
	=
	\mathcal Q_{p,0}^{(n_1,\ldots,n_l)}(r)\,
	a_{00}^{(n_1,\ldots,n_l)}.
\end{equation}
\end{widetext}
The radial integration entering Eq. \ref{Fexd} selects the component at $k=0$
\begin{align}
	a_{00}^{(n_1,\ldots,n_l)}
	\widetilde{\mathcal Q}_{p,0}^{(n_1,\ldots,n_l)}(0)
	&=
	\Omega_d
	\int_0^\infty dr\,
	r^{d-1}
	\nonumber\\
	&\quad\times
	a_{00}^{(n_1,\ldots,n_l)}
	\mathcal Q_{p,0}^{(n_1,\ldots,n_l)}(r).
	\label{Qtilde}
\end{align}
Since $w_p(\mathbf r)=v_p(\hat r)q_p(r)$, the radial coefficient
$\mathcal Q_{p,0}^{(n_1,\ldots,n_l)}(r)$ is obtained from successive convolutions involving only $\Sigma_0(r)$,
and
$-q_p(r)$.
Because
$\beta>0$,
$\Sigma_0(r)>0$, see Appendix \ref{appendixII}-(i),
and
$-q_p(r)\ge0$,
every radial contribution entering these convolutions is positive. Therefore,
$\mathcal Q_{p,0}^{(n_1,\ldots,n_l)}(r)>0$,
which, by Eq.~\ref{Qtilde}, implies $\widetilde{\mathcal Q}_{p,0}^{(n_1,\ldots,n_l)}(0)>0$.

The quantity
$
\langle
\widetilde{\mathcal C}_p^{(n_1,\ldots,n_l)}
(\mathbf 0,\hat d_i,\hat d_j)
\rangle_{\hat k_{ij}}
$
is the isotropic component of the zero-wavevector convolution product of $\tilde{\mathcal{C}}_p^{(n_{\alpha})}(\boldsymbol q,\hat d_i,\hat d_j,\beta)$ in Eq. \ref{Cn_iso}. Each elementary factor entering the convolution has $\hat q$-component
$
[v_p(\hat q,\hat d_i,\hat d_j)]^{n_\alpha}
$.
Expanding each factor in spherical harmonics,
\begin{equation}
	[v_p(\hat q,\hat d_i,\hat d_j)]^{n_\alpha}
	=
	\sum_{\ell,\mu}
	a_{\ell\mu}^{(n_\alpha)}
	(\hat d_i,\hat d_j)
	Y_{\ell\mu}(\hat q),
\end{equation}
the isotropic coefficient is related to
\begin{equation}
	\mu_{n_\alpha}
	=
	\langle
	[v_p(\hat q,\hat d_i,\hat d_j)]^{n_\alpha}
	\rangle_{\hat q} \label{mun}
\end{equation}
by
$
a_{00}^{(n_\alpha)}
=
\sqrt{\Omega_d}\,\mu_{n_\alpha},
$
with
$
Y_{00}=1/\sqrt{\Omega_d}.
$
Assuming that the dominant isotropic contribution is directly given by the $\ell=0$ projection of products of $v_p$, one obtains
\begin{equation}
	\langle
	\widetilde{\mathcal C}_p^{(n_1,\ldots,n_l)}
	(\mathbf 0,\hat d_i,\hat d_j)
	\rangle_{\hat k_{ij}}
	=
	A_{n_1,\ldots,n_l}
	\prod_{\alpha=1}^{l}
	\mu_{n_\alpha}(\hat d_i\cdot\hat d_j), \label{propmun}
\end{equation}
Here
$
A_{n_1,\ldots,n_l}
=
(\sqrt{\Omega_d})^l
\widetilde{\mathcal Q}_{p,0}^{(n_1,\ldots,n_l)}(0)
$. The positivity of $A_{n_1,\ldots,n_l}$ follows from the positivity of
$\widetilde{\mathcal Q}_{p,0}^{(n_1,\ldots,n_l)}(0)$. 
Contributions to
$
\langle
\widetilde{\mathcal C}_p^{(n_1,\ldots,n_l)}
(\mathbf 0,\hat d_i,\hat d_j)
\rangle_{\hat k_{ij}}
$
generated by the coupling of higher-order harmonics in the expansion of powers of $v_p$ are neglected here. The direct isotropic contribution is assumed to be dominant, thus considerably simplifying the analysis.
As shown in Appendix \ref{appendixII}-(ii), for $d \geq 3$,
\begin{eqnarray}
	\max[\mu_n(\hat{d}_i \cdot \hat{d}_j)]=\mu_n(1),
\end{eqnarray}
which under Eq. \ref{propmun} and the positivity of $A_{n_1,\ldots,n_l}$ implies that $\langle
\mathcal C_p^{(n_1,\ldots,n_l)}
(\mathbf r_{ij},\hat d_i,\hat d_j)
\rangle_{\hat r_{ij}}$ is also maximized for $\hat d_i \cdot \hat d_j=1$.

Since $\langle \mathcal{C}^1\rangle_{\hat r_{ij}}=0$, the averaging over $\hat{r}_{ij}$ introduce, similarly to the case $d \rightarrow \infty$, a screening of the dipolar interaction, making it shorter range.

The excess free energy in Eq. \ref{FdRd} is minimized at $\hat{d}_i \cdot \hat{d}_j = 1$ provided the constant $C_W^{d}$, which determines the sign of the reaction-field term, is positive.
The sign of $C_W^{d}$ is fixed by the linear term in the Taylor expansion of $e^{-\beta W_d}$ around $w_p=0$, with $W_d = -\beta^{-1}\mathcal{C}_p$.
Since the linear term in the expansion of $e^{\mathcal C_p}$ is $\mathcal C_p$, inspection of Eq. \ref{Cn} shows that the only contribution to $\mathcal C_p$ linear in $w_p$ is $\mathcal C_p^{(1)}$. Taking the functional derivative of Eq. \ref{C1} with respect to $w_p$ yields the product of two isotropic kernels $\Sigma_0$, each depending on different spatial coordinates. The resulting contribution is therefore quadratic in the radial integral of $\Sigma_0$ and hence positive.
Consequently, both the intrinsic bulk and reaction-field contributions in Eq. \ref{FdRd} favor dipole alignment.

Beyond the exponential approximation Eq. \ref{expapprox}, different representations of the pair density can be employed, raising the question of whether the ferroelectric character of the excess–free–energy minimum is preserved. 
In the high-temperature approximation (HTA), corresponding to the first-order truncation of the lambda-expansion \cite{Hansen}, the term $e^{-\beta W_d}-1$ in Eq. \ref{FdRd} is replaced by the bare dipolar interaction $w_p$, and $C_W^d=1$. The intrinsic bulk contribution in Eq. \ref{FdRd} thus vanishes. The reaction-field contribution, i.e. the first term in Eq.~\ref{FdRd}, instead remains ferroelectric-like. In the one-mode or random-phase approximation (RPA), the term $e^{-\beta W_d}-1$ in Eq. \ref{FdRd} is replaced by the square of the bare dipolar interaction. Consequently, the intrinsic bulk contribution to the excess free energy is minimized for $(\hat d_i\cdot\hat d_j)^2=1$, corresponding to quadrupolar (nematic) order. The reaction-field contribution remains instead always ferroelectric-like. 
The HPA, RPA, and exponential approximations, however, represent, in this order, increasingly accurate levels of description, as each includes a progressively larger subset of diagrams in the cluster expansion.

\subsection{A Simplified Mean-Field Expression for the Excess Free Energy} \label{par4_4}
Within mean-field classical DFT, replacing $W_d$ by its orientationally averaged counterpart $\bar W_d$ yields a finite contribution to the intrinsic bulk excess free energy. For the sake of simplicity, even at finite $d$, one may take $\bar W_d = \bar w_p$, obtained by annealed averaging of the dipolar interaction $w_p$ over $\hat r_{ij}$, instead of the renormalized potential introduced in Sec. \ref{par4_2}. At finite $d$, this choice corresponds to assume exact the truncation of the virial expansion at second order. With the mean-field replacement $p \hat{d}_{i(j)} = \boldsymbol{\delta}$, one obtains in three-dimensional space
\begin{eqnarray}
	\bar{w}_{p}^{(3)}(r,\hat{d}_i, \hat{d}_j)=-\frac{1}{\beta}\log{\int_{-1}^1 d \theta_{ij}e^{-\beta p^2 (\frac{l}{r})^{3}\delta^2 [1-3\theta_{ij}^2]}}, \label{wpannealeddelta}
\end{eqnarray}
where $\theta_{ij}=\hat{\delta}\cdot \hat{r}_{ij} \equiv \cos \alpha_{ij}$. 
Since the exponential term enhances configurations with $\theta_{ij}^2$ close to unity, the integral can be approximated by replacing the Boltzmann factor with its argument averaged over a suitable distribution of $\alpha_{ij}$. Choosing $\alpha_{ij}$ uniformly distributed over its period $[0,2\pi]$ yields the distribution
$
f(\theta_{ij})=\frac{1}{\pi \sqrt{1-\theta_{ij}^2}}
$
for $\theta_{ij}\in[-1,1]$, which enhances the contribution of configurations with $\theta_{ij}^2\simeq1$.
One may therefore approximate the effective potential by averaging $w_p$ over a uniform distribution of the angle between $\hat{\delta}$ and $\hat{r}_{ij}$, as in Ref. \cite{Izzo},
\begin{eqnarray}
	\bar{w}_{p}^{(3)}(r,\hat{d}_i, \hat{d}_j) \simeq  p^2 \left( \frac{l}{r} \right)^{3} \delta^2 \left\langle 1 - 3\theta_{ij}^2 \right\rangle_{\alpha_{ij}}, \label{wpapprox_3}
\end{eqnarray}
where $\langle \ \rangle_{\alpha_{ij}}$ denotes an average over a uniform distribution of $\alpha_{ij}$. Even though this approximate expression correctly captures the ferroelectric character of $\bar{w}_p$, it neglects the screening effect introduced by the annealed averaging, which should make $\bar{w}_p$ shorter-ranged than  $w_p$. In Ref. \cite{Izzo}, the emergence of ferroelectricity was related to positional disorder. In light of the results presented in this manuscript, the positional disorder in Ref. \cite{Izzo} is treated as annealed, tailored to the characteristics of a liquid.

\section{Conclusion} \label{cap_concl}
The emergence of ferroelectricity in dipolar liquids, which has been confirmed by several numerical simulation studies \cite{Wei,Weis1,Weis2,Bartke,Camp,Levesque} and more recent experiments on liquid crystals \cite{Chen,Nishikawa,Martelj}, still lacks a solid theoretical foundation. Studies conducted within a mean field approach \cite{Osipov, Osipov1, Cattes, Groh} remain inconclusive. In particular, they do not clarify whether the ferroelectric phase transition in dipolar liquids is a genuine bulk transition or is instead driven by sample-shape dependent surface contribution to the free energy, which do not vanishes in the thermodynamic limit owing to the long-range nature of the dipolar interaction. 
This property underlies the so-called conditional convergence of the dipolar potential. It is noteworthy that numerical simulations performed  using Ewald summation with conducting periodic boundary conditions \cite{Wei, Weis1, Weis2, Camp, Izzo}, for which the surface contribution to the free energy vanishes, nevertheless exhibit behavior consistent with a ferroelectric phase transition. 
To shed light on this issue, one must move beyond the mean field framework and consider pair correlations between dipoles, as well as the associated mean force generating hindered dipole rotation, in the spirit of Kirkwood treatment of the dielectric properties of polar liquids \cite{Kirkwood}. In the Kirkwood approach focus is placed on the mean force acting between nearest-neighbor dipoles, thereby implicitly emphasizing the role of local structure in any analysis of the onset of ferroelectricity in dipolar liquids. Such an approach would not, however, distinguish between a solid and a liquid exhibiting similar local structures. In the present study, a mean reaction-field construction \cite{Onsager, Kirkwood, Leeuv, Neumann, Wei, Stenqvist, Bell} is introduced in order to isolate the intrinsic bulk contribution of dipolar interaction while making vanishing the surface term. 
The existence of a ferroelectric phase transition in dipolar liquids, intrinsic to the bulk, is then demonstrated. This finding is significant, as it establishes that the onset of a ferroelectric phase transition in dipolar liquids, in the thermodynamic limit, may directly affect intrinsic bulk properties, such as the liquid–liquid phase transition in supercooled water or the emergence of related thermodynamic anomalies.

As main result of the present study, the mean force responsible for a hindered dipolar rotation in dipolar liquids favoring dipolar alignment, and the resulting onset of a ferroelectric phase transition, is shown to emerge from the annealed averaging of the pair dipolar interaction over the relative orientation of the intermolecular separation vector between two particles. 
In the limit $d \to \infty$, where the truncation of the virial expansion becomes exact, the dipolar interaction entering the annealed average coincides with the bare dipolar potential, and its annealed counterpart provides an exact description of the free energy. 
In finite dimensions $d \geq 3$, this dipolar interaction is an effective two-body dipolar potential incorporating many-body effects, and corresponds to an approximation of the free energy within the optimized cluster expansion framework \cite{Hansen,Chandler0}. 
Both the infinite- and finite-dimensional cases are analyzed within a classical DFT framework. 
Annealed averaging over the orientation of the intermolecular separation vector, assumed to be uniformly distributed, is well posed provided that, the system remains liquid in its positional degrees of freedom, independently of the dipolar configuration. In particular, the construction does not rely on any separation of characteristic time scales between dipolar and translational degrees of freedom.
The onset of ferroelectricity in dipolar liquids thus emerges as an intrinsic property of the liquid state, rooted in the annealed disorder inherent to the liquid phase. Within this picture, unlike approaches based on short-range local structure, ferroelectric order arises not in spite of the liquid nature of the system, but because of it.
This outcome parallels the absence of frustration in the Sherrington–Kirkpatrick model with annealed disorder, leading to ordered spin phases \cite{Matsuda, Kasai}. This perspective raises further questions, such as whether quenched disorder also supports the onset of a ferroelectric phase transition, or how ferroelectric order, or its absence, interplays with the crystallization of dipolar liquids. The integration of the classical DFT with the replicated liquid theory \cite{Parisi} would allow the present treatment to be extended to the case of dipolar glasses.

Within the present framework, an effective dipolar interaction emerges that is screened by the spatial isotropy enforced by the reference system potential, shorter-ranged than the bare interaction, and ferroelectric-like. While screening induced by the free rotational motion of dipoles, appropriate to the paraelectric phase, has been extensively considered, see e.g. Refs. \cite{Mercer, Frodl}, commonly refereed as Keesom interaction \cite{Keesom}, the screening  originating from the annealed positional disorder characteristic of the liquid state has remained largely unexplored. 

If the dipolar liquid is embedded in a non-conducting medium, both in numerical simulations and in real systems, any macroscopic polarization necessarily generates a depolarization field that depends on the sample shape and on the dielectric constant of the surrounding medium. As a consequence, in the ferroelectric phase, polarized domains can develop \cite{Groh, Griffiths, Campa, Levesque, Wang, Weis3}. The difficulty in stabilizing such a state in a liquid may, speculatively, account for the tendency of deeply supercooled water, possibly in a low-density polarized phase, to crystallize.

As a further result, the present study identifies the pair correlation function in the paraelectric phase as a diagnostic indicator of the propensity toward ferroelectric ordering, which can occur when the correlation function becomes positive.
On this basis, a method is proposed to reconstruct the pair correlation function entering the DFT free-energy functional from the moments of the probability distribution of $\hat{d}_i \cdot \hat{d}_j$ in the paraelectric phase.
The same issue has recently been addressed from a different perspective by integrating supervised machine learning into a classical DFT framework \cite{Sammuller}.
Interestingly, the onset of ferroelectric ordering is also shown to leave a detectable signature on the radial pair correlation function marginalized over the dipolar degrees of freedom, namely in the radial distribution function $g^{(2)}(r)$ measured in numerical simulations after integrating out dipolar variables.

The present study finally places the classical DFT developments of Ref. \cite{Izzo}, which describes the interplay between ferroelectric and liquid–liquid phase transitions in dipolar liquids with reference to supercooled water, on more solid ground.

\begin{acknowledgements}
Support from Ministero Istruzione Universitá Ricerca - Progetti di Rilevante Interesse Nazionale “Deeping our understanding of the Liquid–Liquid transition in supercooled water”, grant 2022JWAF7Y, is acknowledged. The author gratefully acknowledges the referees for insightful comments and valuable suggestions.
\end{acknowledgements}
\appendix
\section{Large-$d$  Polarization, Scaling, and Free Energy} \label{appendixI}
\numberwithin{equation}{section}
In this Appendix, the following results are derived:
\begin{enumerate}
	\renewcommand{\labelenumi}{(\roman{enumi})}
	\item The ansatz $\zeta(\hat d)$ in Eq. \ref{zeta} is shown to
	discriminate, in the limit $d\to\infty$, between paraelectric and
	ferroelectric states.
	\item The entropy loss associated with the breaking of orientational
	isotropy and the emergence of macroscopic polarization is shown to scale
	as $O(d)$ in the large-$d$ limit, and its explicit expression is
	obtained.
	\item The large-$d$ scaling of the molecular dipole moment in Eq. \ref{prescale} and of the radial variable in
	Eq. \ref{double_scaling} is shown to yield an excess free energy
	$\mathcal F$ of $O(d)$.
	\item The change in $\mathcal{F}$ induced by the onset of a polarized state is computed.
	\item The expression of $\bar{f}_{w_p}(\hat{d}_i \cdot \hat{d}_j,h)$ in Eq. \ref{fbardinf} is established. 
\end{enumerate}

\medskip
\noindent (i) Consider the ansatz
\begin{equation}
	\zeta(\hat d)=\frac{1}{Z_d(\boldsymbol{\delta})}
	e^{d\boldsymbol{\delta}\cdot \hat{d}}; \qquad
	Z_d(\boldsymbol{\delta})=\int d\hat d\,
	e^{d\,\boldsymbol{\delta}\cdot \hat d}.
	\label{zetaA}
\end{equation}
In the following, it is shown that, for small $\delta$, $\boldsymbol{\delta}\equiv \delta \hat{\delta} \propto \bar{\mathbf p}$, so that $\boldsymbol{\delta}$ acts as a ferroelectric order parameter. Accordingly, $\zeta(\hat{d})$ describes a paraelectric state for $\boldsymbol{\delta}=0$ and a ferroelectric state for $\boldsymbol{\delta}\neq 0$.

The molecular average dipole moment is defined as
\begin{equation}
	\bar{\mathbf p} \equiv \bar p\int d\hat d\,\zeta(\hat d)\,\hat d. \label{pbar}
\end{equation}
Introducing the variable $u=\hat{d} \cdot \hat{\delta}$, the angular measure becomes
\begin{equation}
	d\hat d = \Omega_{d-1}(1-u^2)^{\frac{d-3}{2}}\,du; \qquad u \in [-1,1] \label{dddu}
\end{equation}
and Eq. \ref{pbar} reads
\begin{align}
	\bar{\mathbf p}
	&=
	\frac{\int_{-1}^{1}du\,u\,
		\Omega_{d-1}(1-u^2)^{\frac{d-3}{2}}e^{d\delta u}}
	{Z_d(\boldsymbol{\delta})}
	\,\hat\delta;
	\nonumber\\
	Z_d(\boldsymbol{\delta})
	&=
	\int_{-1}^{1}du\,
	\Omega_{d-1}(1-u^2)^{\frac{d-3}{2}}e^{d\delta u}.
	\label{pbar_u}
\end{align}
To prove Eq. \ref{dddu}, one considers the integral over the solid angle of a generic function $F(\hat d \cdot \hat{\delta})$,
\begin{equation}
	\int_{\Omega_d} d\hat d\, F(\hat d \cdot \hat{\delta}) = 2\int_{\mathbb{R}^d} d^d x\, \delta(\mathbf{x}^2-1)\, F(\mathbf{x} \cdot \hat{\delta}), \label{Fhatd}
\end{equation}
where $\boldsymbol{x}=x \hat{d}$. The factor $2$ in Eq. \ref{Fhatd} follows from the relationship $\delta(\mathbf{x}^2-1)=\delta(x-1)/(2x)$,. 
Choosing $\hat\delta$ as polar axis in the $\mathbb{R}^d$ space and decomposing
\begin{equation}
	\mathbf{x}= u \ \hat \delta + \mathbf{x}_\perp,
\end{equation}
where $\mathbf{x}_\perp \in \mathbb{R}^{d-1}$, $\mathbf{x}_\perp=\rho \ \hat{n}$, and $\mathbf{x}_\perp \cdot \hat \delta=0$, one obtains
\begin{equation}
	d^d x = du\, d^{d-1}x_\perp
	= du\, \rho^{d-2} d\rho\, d\hat n,
\end{equation}
where $d\hat n$ is the angular measure in the space $\mathbb{R}^{d-1}$.
Hence
\begin{align}
	\int d\hat d\,F(\hat d \cdot \hat{\delta})
	&=
	2\int_{-\infty}^{\infty} du
	\int_0^\infty \rho^{d-2} d\rho
	\nonumber\\
	&\quad\times
	\int_{\Omega_{d-1}} d\hat n\,
	\delta(u^2+\rho^2-1)\,
	F(u).
	\label{Fhatdesp}
\end{align}
Using $\delta(u^2+\rho^2-1)=\delta(\rho-\sqrt{1-u^2})/(2 \rho)$, one finds
\begin{eqnarray}
	2\int_0^\infty \rho^{d-2} d\rho
	\int_{\Omega_{d-1}} d\hat n \ 
	\delta(u^2+\rho^2-1)
	&=&
	\Omega_{d-1}
	(1-u^2)^{\frac{d-3}{2}},
	\nonumber\\
	&&
	u \in [-1,1].
	\label{Fhatint}
\end{eqnarray}
From Eqs. \ref{Fhatdesp}-\ref{Fhatint}, Eq. \ref{dddu} follows. Eq. \ref{pbar_u} is then obtained by noting that $\zeta(\hat d)$ is axially symmetric about $\hat\delta$, so that the angular integral selects only the component of $\hat d$ parallel to $\hat\delta$.

In the limit $d\to\infty$, the integrals in Eq.~\ref{pbar_u} can be evaluated by Laplace’s saddle-point method. To leading order in $d$, both reduce to integrals of the form
\begin{equation}
	I_d(\delta)=\int_{-1}^{1} du\, e^{d \Phi_\delta(u)}\, g(u); 	\qquad \Phi_\delta(u)=\delta u + \frac{1}{2}\log(1-u^2). \label{Id}
\end{equation}
Here $g(u)=\Omega_{d-1} u$ for the numerator in $\bar{\mathbf p}$, and $g(u)=\Omega_{d-1}$ for $Z_d$.
If $\Phi_\delta(u)$ has a unique global maximum at
$u=\bar u(\delta)\in(-1,1)$: $\Phi_\delta'(\bar{u})=0$, $\Phi_\delta''(\bar u)
<0$, then, by Laplace’s
saddle-point method, one obtains to leading order
\begin{equation}
	I_d(\delta)
	\simeq
	g(\bar u)\, e^{d\Phi_\delta(\bar u)}
	\sqrt{\frac{2\pi}{d|\Phi_\delta''(\bar u)|}},
	\label{laplace}
\end{equation}
where primes denote derivatives with respect to $u$.
In the present case 
\begin{equation}
	\Phi_\delta'(u)=\delta-\frac{u}{1-u^2}=0.
\end{equation}
which yields
\begin{equation}
	\bar u(\delta)=\frac{\sqrt{1+4\delta^2}-1}{2\delta}. \label{ubardelta}
\end{equation}
Combining Eqs.~\ref{pbar_u} and \ref{laplace}, one finally obtains
\begin{equation}
	\bar{\boldsymbol{p}}= \bar p \bar u(\delta)\,\hat \delta. \label{pbarfinal}
\end{equation}
For small $\delta$,
\begin{equation}
	\bar{\mathbf p}
	=\bar p 
	\boldsymbol{\delta}
	+O(\delta^3).
\end{equation}
This shows that, close to the paraelectric state, $\boldsymbol{\delta}$ is proportional to the average polarization and can be used as the ferroelectric order parameter.

\medskip
\noindent (ii) The entropic difference between the paraelectric ($\boldsymbol{\delta}=0$) and ferroelectric ($\boldsymbol{\delta}\neq0$) states is computed in the large-$d$ limit.
The dipole orientational distributions in the two states are
\begin{equation}
	\zeta_{\mathbf 0}(\hat d)=\frac{1}{\Omega_{d}},
	\qquad
	\zeta_{\boldsymbol{\delta}}(\hat d)
	=
	\frac{1}{Z_d(\boldsymbol{\delta})}
	e^{d\boldsymbol{\delta}\cdot\hat d},
\end{equation}
where $Z_d(\boldsymbol{\delta})$ is defined in Eq. \ref{zetaA}. The entropy functional is defined as
\begin{equation}
	S[\zeta]
	=
	-N\int d\hat d\,\zeta(\hat d)\log\zeta(\hat d),
	\label{entropy_def}
\end{equation}
and the entropic difference reads
\begin{equation}
	\Delta S(\boldsymbol{\delta})
	\equiv
	S[\zeta_{\boldsymbol{\delta}}]-S[\zeta_{\mathbf 0}]. \label{DeltaS}
\end{equation}
$S[\zeta_{\mathbf 0}]$ is readily obtained,
\begin{equation}
	S[\zeta_{\mathbf 0}]
	=
	N\log\Omega_{d}, \label{S0}
\end{equation}
while
\begin{equation}
	S[\zeta_{\boldsymbol{\delta}}]
	=
	-Nd\,\boldsymbol{\delta}\cdot
	\int d\hat d\,\zeta_{\boldsymbol{\delta}}(\hat d)\hat d
	+
	N\log Z_d(\boldsymbol{\delta}).
\end{equation}
From Eqs. \ref{pbar} and \ref{pbarfinal} it follows 
\begin{equation}
	S[\zeta_{\boldsymbol{\delta}}]
	=	-
	Nd\,\delta\,\bar u(\delta)+
	N\log Z_d(\boldsymbol{\delta})
	. \label{Sdelta} 
\end{equation}
Expressed in terms of $u=\hat{d} \cdot \hat{\delta}$, $Z_d(\boldsymbol{\delta})$, as given in Eq. \ref{pbar_u}, can be evaluated via Laplace’s saddle-point method, as in Eq.\ref{laplace}. From this it follows
\begin{equation}
	\log Z_d(\boldsymbol{\delta})=\log\Omega_{d-1}+d\Phi_\delta(\bar u)+o(d), \label{Zddelta}
\end{equation}
where $\bar u=\bar u(\delta)$ is given in Eq. \ref{ubardelta}.
For $\delta=0$, the saddle point is at $u=0$, and therefore
\begin{equation}
	\log\Omega_{d}
	=
	\log Z_d(\mathbf 0)
	=
	\log\Omega_{d-1}
	+
	o(d). \label{logOmega}
\end{equation}
Combining Eqs. \ref{DeltaS}-\ref{logOmega}, one obtains
\begin{equation}
	\Delta S(\boldsymbol{\delta})
	=
	Nd\left[
	\Phi_\delta(\bar u)
	-
	\delta\bar u
	\right]
	+
	o(d).
\end{equation}
Using the explicit expression of $\Phi_\delta(\bar u)$ in Eq. \ref{Id} yields
\begin{equation}
	\Delta S(\boldsymbol{\delta})
	=
	\frac{Nd}{2}\log\!\big(1-\bar u^2(\delta)\big)
	+
	o(d),
\end{equation}
which implies Eq. \ref{ent_delta}, since $0\le \bar u^2<1$.

\medskip
\noindent (iii) For convenience, the
large-$d$ scaling of $p$ and $r$ are reported again below,
\begin{equation}
	p^2 = d\,\bar p^2; \qquad r = l\left(1 + \frac{\log d + h}{d}\right),
	\qquad h = O(1). \label{scaling_rp}
\end{equation}
In the large-$d$ limit, the prefactor of $w_p$ becomes
\begin{widetext}
\begin{align}
	p^2\left(\frac{l}{r}\right)^d
	=
	d\bar p^2
	e^{
	-d\log\left(1+\frac{\log d+h}{d}\right)
	}
	\xrightarrow[d\to\infty]{}
	d\bar p^2 e^{-(\log d+h)}
	=
	\bar p^2 e^{-h}, \label{p2e-h}
\end{align}
\end{widetext}
which remains finite for $h=O(1)$.
Although Eq. \ref{p2e-h} is strictly valid for $h=O(1)$, it can be
conveniently extended to $h\to+\infty$, since following Eq. \ref{p2e-h} the rescaled dipolar interaction vanishes exponentially in this limit.
The region 
\begin{equation}
	l < r < l\left(1+\frac{\log d}{d}\right); \qquad -\log d < h < 0, \label{rcoreeff}
\end{equation}
is, however, still accessible under the hard-core constraint $r>l$.  Writing
\begin{equation}
	h=-\log d+s,
	\qquad s=O(1),
\end{equation}
one obtains,
\begin{equation}
	p^2\left(\frac{l}{r}\right)^d
	=d\bar p^2
	\left(1+\frac{s}{d}\right)^{-d}
	\xrightarrow[d\to\infty]{} d\bar p^2
	e^{-s}.
\end{equation}
Since the angular factor entering $w_p$ is not positive definite, the
dipolar interaction can become unbounded, possibly inducing instability
in the system. Therefore, the scaling in Eq. \ref{scaling_rp}
corresponds to a well-defined large-$d$ limit if the region in Eq. \ref{rcoreeff} is excluded. This is achieved by introducing, in the large-$d$ limit, the effective core at $l_{eff}=l(1+\frac{\log d}{d})$. 
More generally, any effective core enforcing $h>-h_c$, with 
$h_c=O(1)$, would merely change the finite upper bound of the dipolar interaction, without affecting its functional form in the large-$d$ limit.

From Eq. \ref{scaling_rp}, one obtains
\begin{align}
	dr &= \frac{l}{d}\,dh;
	\nonumber\\
	r^{d-1}
	&=
	l^{d-1}
	\left(1+\frac{\log d + h}{d}\right)^{d-1}
	\xrightarrow[d\to\infty]{}
	l^{d-1}d
	e^{h}.
\end{align}
and finally Eqs. \ref{FIh}-\ref{fbarexp}. These results show that, under the large-$d$ scaling in Eq. \ref{scaling_rp}, the excess free energy $\mathcal F$ scales as $O(d)$, as the entropy does, provided the
reduced density $\rho B_2^{HS}$ is kept finite in the limit
$d\to\infty$.

\medskip
\noindent (iv)  In the following, the excess free-energy difference $\Delta \mathcal F$
between the paraelectric ($\boldsymbol{\delta}=0$) and ferroelectric
($\boldsymbol{\delta}\neq0$) states in the large-$d$
limit is obtained. The excess free energy evaluated on the ansatz in Eq. \ref{zetaA} is decomposed as
\begin{equation}
	\mathcal F[\zeta_{\boldsymbol\delta}]
	=
	\mathcal F[\zeta_{\mathbf0}]
	+
	\Delta\mathcal F(\rho, \boldsymbol\delta),
	\qquad
	\Delta\mathcal F(\rho,\boldsymbol\delta)
	=
	\mathcal F[\zeta_{\boldsymbol\delta}]
	-
	\mathcal F[\zeta_{\mathbf0}].
	\label{F_decompA}
\end{equation}
Using Eqs. \ref{FIh}-\ref{fbarexp} together with the scaling in
Eq. \ref{scaling_rp}, the excess free energy becomes
\begin{widetext}
\begin{align}
	\mathcal F[\zeta]
	&=
	-\frac{N\rho}{\beta}B_2^{HS}d
	\int_{-\infty}^{\infty}dh\,
	e^h e^{-\beta  v_0(h)}
	\int d\hat d_i\,d\hat d_j\,
	\zeta(\hat d_i)\zeta(\hat d_j)\,
	\bar f_{w_p}(\hat d_i \cdot \hat d_j,h), \label{Fexcessdelta} 
\end{align}
\end{widetext}
with $\bar f_{w_p}(\hat d_i \cdot \hat d_j,h)$ given in Eq. \ref{fbarexp}.
Using the ansatz in Eq. \ref{zetaA}, introducing the variables 
\begin{equation}
	u_i=\hat d_i\cdot\hat\delta,
	\qquad
	u_j=\hat d_j\cdot\hat\delta, \label{uijdef}
\end{equation}
and using Eq. \ref{dddu}, in the limit $d \rightarrow \infty$ one finds
\begin{align}
	\zeta(\hat d_i)\zeta(\hat d_j)\,
	d\hat d_i\,d\hat d_j
	&=
	\frac{\Omega_{d-1}^2}{Z_d^2(\delta)}
	e^{
		d\left[
		\delta(u_i+u_j)
		+
		\frac12\log(1-u_i^2)
		\right]
	}
	\nonumber\\
	&\quad\times
	e^{
		d\left[
		\frac12\log(1-u_j^2)
		\right]
	}
	du_i\,du_j .
	\label{angularF}
\end{align}
As shown in point (v) below, the integral in
Eq. \ref{fbarexp} can be expressed in terms of Gaussian random
variables, and the argument of the exponential in Eq. \ref{fbarexp}
remains $O(1)$ in the large-$d$ limit. Therefore, given Eq. \ref{angularF}, the integral in
Eq. \ref{Fexcessdelta} can be evaluated for $d \rightarrow \infty$ by the Laplace saddle-point
method, following the same steps as in Eqs. \ref{Id}-\ref{laplace}, with $\bar{f}_{w_p}$ acting as $g$.
The saddle-point condition reads
\begin{equation}
	u_i=u_j=\bar u, \label{doublesella}
\end{equation}
with $\bar{u}$ in Eq. \ref{ubardelta}. At this point,
\begin{equation}
	\hat{d}_i \cdot \hat{d}_j = \bar{u}^2. \label{ubar2}
\end{equation}
Indeed, upon decomposing
\begin{equation}
	\hat d_{i(j)}
	=
	\bar{u}\hat\delta
	+
	[1-u_{i(j)}^2]^{\frac{1}{2}}\,\hat{e}_{i(j)},
\end{equation}
with $\boldsymbol{e}_{i(j)}\cdot\hat\delta=0$, one obtains
\begin{equation}
	\hat d_i\cdot\hat d_j
	=
	\bar{u}^2
	+
	(1-\bar{u}^2)
	\,(\hat{e}_i\cdot\hat{e}_j).
\end{equation} 
As follows from Eq. \ref{transf}, for $\hat r_{ij}$ uniformly distributed on the unit sphere in $d$ dimensions, the quantity 
$\sqrt d\,(\hat r_{ij}\cdot\hat d_i)$ converges in distribution, in the limit $d\to\infty$, to a centered Gaussian variable with unit variance. In the subspace orthogonal to $\hat\delta$, the same holds for 
$\sqrt{d-1}\,(\hat e_i\cdot\hat e_j)$. Consequently,
$\hat e_i\cdot\hat e_j = O(d^{-1/2})$,
from which Eq. \ref{ubar2} follows.

Eq. \ref{deltaFex} is thus established.

\medskip
\noindent (v)
$\bar{f}_{w_p}(\hat{d}_i \cdot \hat{d}_j,h)$ is defined in Eq. \ref{fbarexp}. Introduce the variables
\begin{equation}
	\theta_i=\hat d_i\cdot\hat r_{ij},
	\qquad
	\theta_j=\hat d_j\cdot\hat r_{ij}.
\end{equation}
To express the angular measure $d\hat r_{ij}$ in terms of
$\theta_i$ and $\theta_j$, consider an orthonormal basis
$(\hat e_1,\hat e_2)$ of the plane spanned by
$\hat d_i$ and $\hat d_j$, 
\begin{equation}
	\hat r_{ij}
	=
	\eta_1\hat e_1
	+
	\eta_2\hat e_2
	+
	\boldsymbol r_\perp .
\end{equation}
The variables $\eta_{1(2)}$ parameterize the components of
$\hat r_{ij}$ in the $(\hat d_i,\hat d_j)$ plane, while
$\boldsymbol r_\perp \in\mathbb R^{d-2}$ represents the component of $\hat{r}_{ij}$ orthogonal to this
plane. 
Applying the same argument used to derive Eq. \ref{dddu}, now resolving
the solid-angle element $d \hat{r}_{ij}$ in terms of the two variables $\eta_1$ and
$\eta_2$ instead of a single variable $u$, one obtains
\begin{equation}
	d\hat r_{ij}
	=
	\Omega_{d-2}
	\left(1-\eta_1^2-\eta_2^2\right)^{\frac{d-4}{2}}
	d\eta_1\,d\eta_2,
	\qquad
	\eta_1^2+\eta_2^2\le 1.
\end{equation}
It remains to determine the relation between the differential elements
$d\eta_1\,d\eta_2$ and $d\theta_i\,d\theta_j$. 
Since $\hat d_i$ and $\hat d_j$ lie in the
$(\hat e_1,\hat e_2)$ plane, the variables
$\mathbf t_{ij}=(\theta_i,\theta_j)$ depend linearly on
$\boldsymbol\eta=(\eta_1,\eta_2)$,
\begin{equation}
	\mathbf t_{ij}
	=
	A\,\boldsymbol\eta,
\end{equation}
with
\begin{equation}
	A=
	\begin{pmatrix}
		\hat d_i\cdot\hat e_1 & \hat d_i\cdot\hat e_2\\
		\hat d_j\cdot\hat e_1 & \hat d_j\cdot\hat e_2
	\end{pmatrix}.
\end{equation}
It follows that 
\begin{equation}
	\eta_1^2+\eta_2^2
	=
	\boldsymbol\eta^T\boldsymbol\eta
	=
	\mathbf t_{ij}\mathbf G^{-1}\mathbf t_{ij}^T,
\end{equation}
where $\mathbf G =A A^T$ is the Gram matrix of $\hat d_i$, $\hat d_j$.
Therefore the integration domain becomes
\begin{equation}
	\mathcal D
	=
	\left\{
	\mathbf t_{ij}=(\theta_i,\theta_j)\in\mathbb R^2:
	\mathbf t_{ij}\mathbf G^{-1}\mathbf t_{ij}^{T}
	\le 1
	\right\}.
\end{equation}
Moreover, for the linear transformation $\mathbf t_{ij}=A\boldsymbol{\eta}$, the integration measure transforms as
\begin{equation}
	d\eta_1\,d\eta_2
	=
	\frac{d\mathbf t_{ij}}{|\det A|}=
	\frac{d\mathbf t_{ij}}{(\det\mathbf G)^{1/2}},
\end{equation}
where the identity $ \det\mathbf G =	\det(AA^T) = (\det A)^2$ has been exploited, and $d\mathbf t_{ij}=d\theta_id\theta_j$ . Eq. \ref{fwp_annealed1} is thus obtained. Introducing then the variable
\begin{equation}
	\tilde{\mathbf t}_{ij}
	=
	\sqrt d\,\mathbf t_{ij},
\end{equation}
one has
\begin{align}
	\left(
	1-\mathbf t_{ij}\mathbf G^{-1}\mathbf t_{ij}^{T}
	\right)^{\frac{d-4}{2}}
	&=
	e^{
		\frac{d-4}{2}
		\log\left(
		1-
		\frac{
			\tilde{\mathbf t}_{ij}\mathbf G^{-1}\tilde{\mathbf t}_{ij}^{T}
		}{d}
		\right)
	}
	\\
	&\xrightarrow[d\to\infty]{}
	e^{
		-\frac12
		\tilde{\mathbf t}_{ij}\mathbf G^{-1}\tilde{\mathbf t}_{ij}^{T}
	}.
\end{align}
Moreover, using
$\Omega_d=\frac{2\pi^{d/2}}{\Gamma(d/2)}$, where $\Gamma(x)$ denotes
the Gamma function satisfying
$\Gamma(x+1)=x\Gamma(x)$, one obtains
\begin{align}
	\frac{\Omega_{d-2}}{\Omega_d}
	=
	\frac{d-2}{2\pi}.
\end{align}
Since
\begin{equation}
	d\tilde{\mathbf t}_{ij}
	=d \ 
	d\mathbf {t}_{ij},
\end{equation}
Eq. \ref{transf} is thus retrieved. The large-$d$ scaling in Eq. \ref{scaling_rp} finally leads to Eq. \ref{fbardinf}. Eq. \eqref{fbardinf_1} shows that $\bar f(r,\hat r_{ij},\hat d_i,\hat d_j)$ can be rewritten in terms of the
moment-generating function $M_{\tilde Z}(t)$ of the random variable
$\tilde Z=\tilde\theta_i\tilde\theta_j-\hat d_i\cdot\hat d_j$:
\begin{equation}
	M_{\tilde Z}(t)
	=
	\langle e^{t(\tilde\theta_i\tilde\theta_j-\hat d_i\cdot\hat d_j)}\rangle
	=
	e^{-t\hat d_i\cdot\hat d_j}
	\langle e^{t\tilde\theta_i\tilde\theta_j}\rangle .
\end{equation}
Since \((\tilde\theta_i,\tilde\theta_j)\) is distributed as a centered bivariate Gaussian with covariance matrix \(\mathbf G\), one can rewrite
\[
-\frac12 \tilde{\mathbf t}_{ij}^{T}\mathbf G^{-1}\tilde{\mathbf t}_{ij}
+t\,\tilde\theta_i\tilde\theta_j
=
-\frac12
\tilde{\mathbf t}_{ij}^{T}
\left(
\mathbf G^{-1}-t\tilde{\mathbf I}
\right)
\tilde{\mathbf t}_{ij},
\]
where $\tilde{\mathbf I}$ is the symmetric matrix with vanishing diagonal entries and unit off-diagonal entries. Performing the Gaussian integral then gives,
\begin{equation}
	\langle e^{t\tilde\theta_i\tilde\theta_j}\rangle
	=
	\frac{1}{
		\sqrt{
			\det\mathbf G\,
			\det(\mathbf G^{-1}-t\tilde{\mathbf I})
	}},
\end{equation}
Using
\begin{equation}
	\det\mathbf G\,
	\det(\mathbf G^{-1}-t\tilde{\mathbf I})
	=
	1-2tq-t^2(1-q^2),
\end{equation}
one finally obtains Eq. \eqref{MGF}.

\section{Properties of $\Sigma_0$ and $\mu_n$ at Finite Dimension} \label{appendixII}
This Appendix establishes the following results:
\begin{enumerate}
	\renewcommand{\labelenumi}{(\roman{enumi})}
	\item The positivity of the hypervertex function $\Sigma_0(r)$ entering $\mathcal C_p^{(n)}$ for dipolar interaction.
	\item The quantities $\mu_n$ defined in Eq.~\ref{mun} are maximal when
	$\hat d_i \cdot \hat d_j=1$.
\end{enumerate}

\medskip
\noindent (i)
It is convenient to rewrite the hypervertex function in terms of the pair correlation function $g_0(\mathbf r_i,\mathbf r_j)$ rather than $h_0(\mathbf r_i,\mathbf r_j)$. Eq. \ref{sigma} then becomes
\begin{eqnarray}
	\Sigma_0(\mathbf r_i,\mathbf r_j)
	=
	\rho\delta(\mathbf r_i-\mathbf r_j)
	-\rho^2
	+\rho^2 g_0(|\mathbf r_i-\mathbf r_j|).
	\label{sigmag}
\end{eqnarray}
Consider the generic $\mathcal C_p^{(n)}$ obtained trough generalization of Eq. \ref{C1},
\begin{widetext}
\begin{multline}
	\rho^2 \mathcal C_p^{(n)}
	=
	-\beta \int
	d\mathbf r_1\cdots d\mathbf r_{n+1}\,
	\big[
	\rho\delta(\mathbf r_i-\mathbf r_1)
	-\rho^2
	+\rho^2 g_0(|\mathbf r_i-\mathbf r_1|)
	\big]
	\\
	\times
	w_p(\mathbf r_1-\mathbf r_2,\hat d_1,\hat d_2)
	\Sigma_0(\mathbf r_2,\mathbf r_3)
	\cdots
	w_p(\mathbf r_n-\mathbf r_{n+1}, \hat d_n,\hat d_{n+1})
	\Sigma_0(\mathbf r_{n+1},\mathbf r_j).
\end{multline}
\end{widetext}
If the term $-\rho^2$ is selected from the first hypervertex $\Sigma_0(\mathbf r_i,\mathbf r_1)=\big[
\rho\delta(\mathbf r_i-\mathbf r_1)
-\rho^2
+\rho^2 g_0(|\mathbf r_i-\mathbf r_1|)
\big]$, the integration over $\mathbf r_1$ factorizes and produces a term proportional to
\begin{equation}
	\int d\mathbf r_1\,
	w_p(\mathbf r_1-\mathbf r_2,\hat d_1,\hat d_2)=0, \label{vp0}
\end{equation}
because $\langle v_p(\hat r_{ij},\hat d_i,\hat d_j)\rangle_{\hat r_{ij}}=0$. The same holds whenever the term $-\rho^2$ is selected from any hypervertex $\Sigma_0(\mathbf r_{m-1},\mathbf r_m)$ in the convolution chain: the integration over $\mathbf r_{m-1}$ factorizes and yields an integral proportional to
\begin{equation}
	\int d\mathbf r_{m-1}\,
	w_p(\mathbf r_{m-1}-\mathbf r_m,\hat d_{m-1},\hat d_m),
\end{equation}
which again vanishes by Eq.~\ref{vp0}. Therefore every contribution containing the constant term $-\rho^2$ vanishes identically at any order in $\mathcal C_p^{(n)}$, $\forall n$.
The hypervertex can thus be replaced by the effective isotropic function
\begin{eqnarray}
	\tilde\Sigma_0(r)
	=
	\rho\,\frac{\delta(r)}{\Omega_d r^{d-1}}
	+
	\rho^2 g_0(r),
	\label{sigmatilde}
\end{eqnarray}
where the first term is the radial representation of the $d$-dimensional Dirac delta.
In the nontrivial case where $g_0(r)$ is not identically zero, one has $g_0(r)>0$ on a finite interval, and consequently $\tilde\Sigma_0(r)$ is positive. 

\medskip
\noindent (ii) 
In the following it is shown that the quantities $\mu_n$ in Eq. \ref{mun} are maximal when
$
\hat d_i \cdot \hat d_j=1$.  Note that in the following the integration variable has been changed from $\hat q$ in Eq. \ref{mun} to $\hat r_{ij}$ for notational convenience only, since it is indeed a dummy variable.
Defining
\begin{equation}
	v_p(\hat r_{ij},\hat d_i, \hat d_j)
	=
	\hat r_{ij}^{\mathrm T}
	A(\hat d_i, \hat d_j)
	\hat r_{ij},
\end{equation}
with $A(\hat d_i,\hat d_j)$ the real symmetric $d\times d$ matrix
\begin{equation}
	A(\hat d_i, \hat d_j)
	=
	\frac d2
	\left(
	\hat d_i\hat d_j^{\mathrm T}
	+
	\hat d_j\hat d_i^{\mathrm T}
	\right)
	-
	\hat d_i \cdot \hat d_j \,\mathbf I ,
\end{equation}
one can write, up to the positive normalization factor $\frac{1}{\Omega_d}$,
\begin{equation}
	\mu_n
	=
	\int
	d\hat r_{ij}\,
	\left(
	\hat r_{ij}^{\mathrm T}
	A(\hat d_i, \hat d_j)
	\hat r_{ij}
	\right)^n. \label{mun_A}
\end{equation}
Consider the generalized binomial expansion
\begin{equation}
	(1-x)^{-d/2}
	=
	\sum_{n=0}^{\infty}
	\frac{(d/2)_n}{n!}x^n .
	\label{genbin}
\end{equation}
Here
\begin{equation}
	\frac{(d/2)_n}{n!}
	=
	\frac{\Gamma(d/2+n)}
	{\Gamma(d/2)\Gamma(n+1)},
\end{equation}
where $\Gamma(x)$ is the Euler Gamma function.
It is then convenient to introduce the auxiliary generating function
\begin{equation}
	H_{\hat d_i\cdot\hat d_j}(z)
	=
	\sum_{n=0}^{\infty}
	\frac{(d/2)_n}{n!}\,
	\mu_n\,
	z^n . \label{H}
\end{equation}
The quantities $\mu_n$ are therefore directly encoded in the Taylor coefficients of
$
H_{\hat d_i\cdot\hat d_j}(z)
$,
up to the positive factors $(d/2)_n/n!$.
Using Eq. \ref{mun_A}, exchanging the order of summation and integration, and applying Eq. \ref{genbin} with
$
x=zv_p(\hat r_{ij}, \hat d_i, \hat d_j),
$
one obtains
\begin{equation}
	H_{\hat d_i\cdot\hat d_j}(z)
	=
	\int
	d\hat r_{ij}\,
	\left[
	1-z
	\hat r_{ij}^{\mathrm T}
	A(\hat d_i,\hat d_j)
	\hat r_{ij}
	\right]^{-d/2}.
\end{equation}
The function $H_{\hat d_i \cdot \hat d_j}(z)$ is introduced because studying the sign and monotonicity of $\mu_n$ is equivalent to studying its Taylor coefficients.

To evaluate $H_{\hat d_i\cdot\hat d_j}(z)$, it is convenient to work in the eigenbasis of
$
A(\hat d_i,\hat d_j)
$.
Since it is a real symmetric matrix, it can be diagonalized by an orthogonal transformation. In its eigenbasis,
\begin{equation}
	\hat r_{ij}^{\mathrm T}
	A(\hat d_i,\hat d_j)
	\hat r_{ij}
	=
	\sum_{\alpha=1}^{d}
	\lambda_\alpha  r_{ij,\alpha}^2,
\end{equation}
where $\lambda_\alpha$ are the eigenvalues of $A(\hat d_i,\hat d_j)$, $\alpha=1,\ldots,d$ labels the corresponding orthonormal eigenvectors $\boldsymbol{e}_\alpha$, and $ r_{ij,\alpha}=\hat r_{ij}\cdot \boldsymbol{e}_\alpha$.
The eigenvalues of $A(\hat d_i,\hat d_j)$ are
\begin{equation}
	\lambda_\pm
	=
	\frac{(d-2)(\hat d_i\cdot\hat d_j)\pm d}{2},
	\qquad
	\lambda_0=-(\hat d_i\cdot\hat d_j), \label{eigenvalues}
\end{equation}
where $\lambda_0$ has multiplicity $d-2$. To derive Eq, \ref{eigenvalues}, one can choose an orthonormal basis
\begin{equation}
	\hat d_i=\boldsymbol{e}_1,
	\qquad
	\hat d_j=
	(\hat d_i\cdot\hat d_j)\boldsymbol{e}_1
	+
	[1-(\hat d_i\cdot\hat d_j)^2]^{\frac{1}{2}}\,\boldsymbol{e}_2 .
\end{equation}
On the plane spanned by $\boldsymbol e_1$ and $\boldsymbol e_2$, the restriction of $A(\hat d_i,\hat d_j)$ takes the form
\begin{equation}
	\mathcal{A}(\hat d_i,\hat d_j)
	=
	\begin{pmatrix}
		(d-1)(\hat d_i\cdot\hat d_j)
		&
		\frac d2[1-(\hat d_i\cdot\hat d_j)^2]^{\frac{1}{2}}
		\\
		\frac d2[1-(\hat d_i\cdot\hat d_j)^2]^{\frac{1}{2}}
		&
		-(\hat d_i\cdot\hat d_j) \label{A2x2}
	\end{pmatrix}.
\end{equation}
On the orthogonal complement, one has
\begin{equation}
	\mathcal{A}_{\perp}(\hat d_i,\hat d_j)=-(\hat d_i\cdot\hat d_j)\mathbf I ,
\end{equation}
so that $\lambda_0=-(\hat d_i\cdot\hat d_j)$ is an eigenvalue with multiplicity $d-2$.
The remaining two eigenvalues are those of the restricted matrix $\mathcal A(\hat d_i,\hat d_j)$. In the diagonal basis of $A(\hat d_i,\hat d_j)$, 
\begin{equation}
	H_{\hat d_i\cdot\hat d_j}(z)
	=
	\int
	d\hat r_{ij}\,
	(
	1-z
	\sum_{\alpha=1}^{d}
	\lambda_\alpha
	r_{ij,\alpha}^2
	)^{-d/2},
\end{equation}
The uniform measure on the unit sphere can equivalently be generated from Gaussian variables. Let $\mathbf g=g\hat g \in\mathbb R^d$
be a vector distributed according to the normalized Gaussian measure
\begin{equation}
	dP(\mathbf g)
	=
	\frac{
		d\mathbf g\,
		e^{- g^2/2}
	}{
		\int_{\mathbb R^d}d\mathbf g\,
		e^{- g^2/2}
	}.
\end{equation}
Then the unit vector $\hat g =\frac{\mathbf g}{|\mathbf g|}$ is uniformly distributed on the $d$-dimensional unit sphere. Therefore,
\begin{equation}
	H_{\hat d_i\cdot\hat d_j}(z)
	=
	\frac{
		\int_{\mathbb R^d}d\mathbf g\,
		e^{- g^2/2}
		\left(
		1-z
		\sum_{\alpha=1}^{d}
		\lambda_\alpha
		\frac{g_\alpha^2}{ g^2}
		\right)^{-d/2}
	}{
		\int_{\mathbb R^d}d\mathbf g\,
		e^{- g^2/2}
	}.
\end{equation}
Using spherical coordinates with $d\mathbf g=g^{d-1}dg\,d\hat g$ and $g_\alpha/g=\hat g_\alpha$, the radial integral factorizes both in the numerator and in the denominator and cancels, yielding
\begin{equation}
	H_{\hat d_i\cdot\hat d_j}(z)
	=
	\frac{1}{\Omega_d}
	\int d\hat g\,
	(
	1-z
	\sum_{\alpha=1}^{d}
	\lambda_\alpha
	\hat g_\alpha^2
	)^{-d/2}.
\end{equation}
Define
\begin{equation}
	I(z)
	=
	\int_{\mathbb R^d}d\mathbf g\,
	e^{
		-\frac12
		\sum_{\alpha=1}^{d}
		(1-z\lambda_\alpha)g_\alpha^2
	}.
\end{equation}
Using spherical coordinates, it transforms into
\begin{equation}
	I(z)
	=
	\int d\hat g
	\int_0^\infty dg\,
	g^{d-1}
	e^{
		-\frac{g^2}{2}
		(
		1-z
		\sum_{\alpha=1}^{d}
		\lambda_\alpha
		\hat g_\alpha^2
		)}.
\end{equation}
Since
\begin{equation}
	\int_0^\infty dg\,
	g^{d-1}
	e^{-\frac12 a g^2}
	=
	2^{\frac d2-1}
	\Gamma\left(\frac d2\right)
	a^{-d/2},
\end{equation}
it follows that
\begin{equation}
	I(z)
	=
	2^{\frac d2-1}
	\Gamma\left(\frac d2\right)
	\int d\hat g\,
	(
	1-z
	\sum_{\alpha=1}^{d}
	\lambda_\alpha
	\hat g_\alpha^2
	)^{-d/2}.
\end{equation}
Computing $I(z)$ at $z=0$, one obtains
\begin{equation}
	H_{\hat d_i\cdot\hat d_j}(z)
	=
	\frac{I(z)}{I(0)}.
\end{equation}
On the other hand,
\begin{equation}
	I(z)
	=
	\prod_{\alpha=1}^{d}
	\int_{-\infty}^{\infty}dg_\alpha\,
	e^{-\frac12(1-z\lambda_\alpha)g_\alpha^2}.
\end{equation}
Using
\begin{equation}
	\int_{-\infty}^{\infty}dg\,
	e^{-\frac12 a g^2}
	=
	\sqrt{\frac{2\pi}{a}},
\end{equation}
thus one finally obtains
\begin{equation}
	H_{\hat d_i\cdot\hat d_j}(z)
	=
	\prod_{\alpha=1}^{d}
	(1-z\lambda_\alpha)^{-1/2}. \label{H_eigen}
\end{equation}
With $q=\hat d_i\cdot\hat d_j$, from Eq. \ref{H_eigen},
\begin{equation}
	\log H_q(z)
	=
	\sum_{m=1}^{\infty}
	\frac{p_m(q)}{2m}z^m;
	\qquad
	p_m(q)
	=
	\sum_{\alpha=1}^{d}\lambda_\alpha^m .
\end{equation}
Using the spectrum of $A(\hat d_i,\hat d_j)$, see Eq.~\ref{eigenvalues},
\begin{equation}
	p_m(q)=
	\left[
	\frac{d-2}{2}q+\frac{d}{2}
	\right]^m
	+
	\left[
	\frac{d-2}{2}q-\frac{d}{2}
	\right]^m
	+(d-2)(-q)^m .
\end{equation}
Since $A(\hat d_i,\hat d_j)$ is traceless, $p_1(q)=0$.
As follows immediately from the explicit form of the eigenvalues,
$
p_m(-q)=(-1)^m p_m(q)$.
It is therefore sufficient to analyze the sign and monotonicity properties for
$
q>0
$.
For $d\ge3$ and
$
0<q\le1,
$
one has
\begin{equation}
	p_m(q)>0,
	\qquad
	m\ge2 .
	\label{pmpos}
\end{equation}
For even $m$, this is immediate. For odd $m\ge3$, using the binomial expansion, one obtains
\begin{widetext}
\begin{equation}
	\left(
	\frac d2
	+
	\frac{d-2}{2}q
	\right)^m
	-
	\left(
	\frac d2
	-
	\frac{d-2}{2}q
	\right)^m
	=
	2
	\sum_{\ell\ {\rm odd}}
	\binom{m}{\ell}
	\left(
	\frac d2
	\right)^{m-\ell}
	\left(
	\frac{d-2}{2}q
	\right)^\ell .
\end{equation}
\end{widetext}
All terms in the sum are positive. Keeping only the first term gives
$
m(d-2)
\left(
\frac d2
\right)^{m-1}
q
$.
Since $m\ge3$, $d\ge3$, and
$
0<q\le1,
$
one has
$
m
\left(
\frac d2
\right)^{m-1}
q
>
q^m
$.
Therefore Eq. \ref{pmpos} follows. Moreover, by the same argument, one obtains
\begin{equation}
	\frac{d p_m(q)}{dq}
	>0,
	\qquad
	m\ge2,
	\quad
	q>0 .
\end{equation}
Thus all nonzero Taylor coefficients of
$
\log H_q(z)
$
and of
$
\partial_q \log H_q(z)
$
are positive for
$
q>0.
$
Since
\begin{equation}
	H_q(z)
	=
	e^{\!
	\log H_q(z)
	}; \ \ \ 
	\partial_q H_q(z)
	=
	H_q(z)
	\,
	\partial_q \log H_q(z),
\end{equation}
the Taylor coefficients of
$
H_q(z)
$
and of
$
\partial_q H_q(z)
$
are themselves positive.
Comparing with Eq.~\ref{H}
and considering that
$
(d/2)_n/n!>0,
$
it follows
\begin{equation}
	\mu_n(q)>0,
	\qquad
	\frac{d\mu_n(q)}{dq}>0,
	\label{munposeder}
\end{equation}
for
$
n\ge2,
$
$
d\ge3,
$
and
$
q>0
$.
Moreover, since
$
p_m(-q)
=
(-1)^m
p_m(q)
$,
the moments inherit the same parity structure:
$
\mu_n
$
is even for even $n$ and odd for odd $n$.
Furthermore, for even $n$,
$
\partial_q\mu_n
$
is odd, whereas for odd $n$ it is even.
Therefore, Eq.~\ref{munposeder}
implies that, for even $n$, $\mu_n$ increases with
$
|q|
$,
while for odd $n$ it is monotonically increasing on the whole interval
$
[-1,1]
$.
In particular,
\begin{equation}
	\mu_n(q)
	\le
	\mu_n(1),
\end{equation}
so that $\mu_n$ are maximal for
$
q=1
$.

\bibliographystyle{apsrev4-1}
\bibliography{achemso-demo}
	
\end{document}